\documentclass[twocolumn,aps,pra,superscriptaddress]{revtex4}
\usepackage[latin9]{inputenc}
\usepackage{color}
\usepackage{bm}
\usepackage{amsmath}
\usepackage{amssymb}
\usepackage[bookmarks=false,
 breaklinks=false,pdfborder={0 0 1},backref=section,colorlinks=true]
 {hyperref}
\hypersetup{
 linkcolor=blue,citecolor=blue,urlcolor=blue}

\makeatletter
\@ifundefined{textcolor}{}
{%
 \definecolor{BLACK}{gray}{0}
 \definecolor{WHITE}{gray}{1}
 \definecolor{RED}{rgb}{1,0,0}
 \definecolor{GREEN}{rgb}{0,1,0}
 \definecolor{BLUE}{rgb}{0,0,1}
 \definecolor{CYAN}{cmyk}{1,0,0,0}
 \definecolor{MAGENTA}{cmyk}{0,1,0,0}
 \definecolor{YELLOW}{cmyk}{0,0,1,0}
}

\usepackage{bm}
\usepackage{color}
\usepackage{braket}
\usepackage{subfigure}\usepackage{epsfig}\usepackage{amsfonts}\usepackage{mathrsfs}\usepackage[toc,page,title,titletoc,header]{appendix}

\@ifundefined{textcolor}{}{%
 \definecolor{BLACK}{gray}{0}
 \definecolor{WHITE}{gray}{1}
 \definecolor{RED}{rgb}{1,0,0}
 \definecolor{GREEN}{rgb}{0,1,0}
 \definecolor{BLUE}{rgb}{0,0,1}
 \definecolor{CYAN}{cmyk}{1,0,0,0}
 \definecolor{MAGENTA}{cmyk}{0,1,0,0}
 \definecolor{YELLOW}{cmyk}{0,0,1,0}
}

\makeatother

\begin{document}
\title{  Herman-Kluk-Like Semi-Classical Initial-Value Representation for   Boltzmann Operator}
\author{Binhao Wang}
\affiliation{School of Physics, Renmin University of China, Beijing, 100872, China}
\author{Fan Yang}
\affiliation{Hefei National Laboratory, Hefei 230088, China}
\author{Chen Xu}
\affiliation{School of Physics, Renmin University of China, Beijing, 100872, China}
\author{Peng Zhang}
\email{pengzhang@ruc.edu.cn}

\affiliation{School of Physics, Renmin University of China, Beijing, 100872, China}
\affiliation{Key Laboratory of Quantum State Construction and Manipulation (Ministry
of Education), Renmin University of China, Beijing, 100872, China}
\date{\today}
\begin{abstract}
The coherent-state initial-value representation (IVR) for the semi-classical real-time propagator of a quantum system, developed by Herman and Kluk (HK), is widely used in computational studies of chemical dynamics. On the other hand, the Boltzmann operator 
$
e^{-\hat{H}/(k_B T)},
$
with \( \hat{H} \), \( k_B \), and \( T \) representing the Hamiltonian, Boltzmann constant, and temperature, respectively, plays a crucial role in chemical physics and other branches of quantum physics.
One might naturally assume that a semi-classical IVR for the matrix element of this operator in the coordinate representation (i.e., $
\langle \tilde{\bm{x}} | e^{-\hat{H}/(k_B T)} | \bm{x} \rangle
$, or the imaginary-time propagator) 
could be derived via a straightforward ``real-time $\rightarrow$ imaginary-time transformation'' from the HK IVR of the real-time propagator. However, this is not the case, as such a transformation results in a divergence in the high-temperature limit (\(T \rightarrow \infty\)).
In this work, we solve this problem and develop a reasonable HK-like semi-classical IVR for 
$
\langle \tilde{\bm{x}} | e^{-\hat{H}/(k_B T)} | \bm{x} \rangle,
$
specifically for systems  where either the gradient of the potential energy (i.e., the force intensity) has a finite upper bound, or the potential becomes harmonic in the long-range limit. The integrand in this IVR is a real Gaussian function of the positions \( \bm{x} \) and \( \tilde{\bm{x}} \), which facilitates its application to realistic problems. Our HK-like IVR is exact for free particles and harmonic oscillators, and its effectiveness for other systems is demonstrated through numerical examples.

\end{abstract}
\maketitle

\section{Introduction}
\label{introduction}

In 1928, van Vleck derived the semi-classical real-time propagator (the van Vleck propagator) for a quantum system in the limit as $\hbar\rightarrow0$ \cite{vv}. The phase of the van Vleck propagator, with respect to a given initial and final time and position, is determined by the action of the corresponding classical trajectory. The van Vleck propagator reveals the intrinsic connection between quantum and classical mechanics, having had a significant impact on  quantum physics \cite{sc1,sc2}.

However, deriving the van Vleck propagator requires solving the classical Hamiltonian equations with fixed initial and final positions. This is not convenient for numerical calculations of realistic systems due to the root-finding problem. Consequently, the van Vleck propagator is not well-suited for numerical studies of atom or molecule systems. To overcome this problem, many authors \cite{preHK1, preHK2, preHK3, preHK4, preHK1b, preHK5, HK} have attempted to develop initial-value representations (IVRs) for semi-classical real-time propagators. In the IVRs, these propagators are expressed as functionals of classical trajectories, determined by the given initial position and momentum, which can be easily derived numerically via standard algorithms, such as Runge-Kutta. A highly influential IVR for the semi-classical real-time propagator was derived by Herman and Kluk (HK) in 1984 \cite{HK}, based on coherent states. Over the past forty years, the HK representation has been widely used in studies of chemical dynamics, and has proven to be a valuable tool for exploring the quantum effects of nuclear motion \cite{R1, R2}. Moreover, in 2006, Kay provided a rigorous derivation of the HK representation from the Schr\"odinger equation, demonstrating that this representation is the leading term in an asymptotic expansion of the quantum propagator in powers of 
$\hbar$~\cite{Kay}.

In addition to the real-time propagator, 
the {\it imaginary-time} propagator  $
\langle \tilde{\bm{x}} | e^{-\hat{H}/(k_B T)} | \bm{x} \rangle
$, which is
the matrix element  of the
Boltzmann operator $e^{-\hat{H}/(k_{B}T)}$  in the coordinate representation, also play a crucial role in quantum physics and chemistry. Here, \( k_B \) and \( T \) are the Boltzmann constant and temperature, respectively. 
 In 1971, Miller derived the semi-classical imaginary-time propagator \cite{Miller1971}, through a direct \( t \rightarrow -i\tau \) transformation of the van Vleck propagator, where \( t \) is the real time and \( \tau = \hbar/(k_B T) \). Similar to the van Vleck propagator, the result obtained by Miller in Ref.~\cite{Miller1971} is determined by the classical trajectory with an inverted potential, with respect to fixed initial and final positions.

An semi-classical IVR for the imaginary-time propagator, which is similar to the HK representation for the real-time one, would clearly be very useful for numerical studies in atom physics, molecule physics and chemical physics. Intuitively speaking, one might expect to obtain such an IVR through the aforementioned \( t \rightarrow -i\tau \) transformation from the HK representation of the real-time propagator \cite{Miller}. However, as pointed out by Yan, Liu, and Shao in Ref.~\cite{Yan and Shao}, this is not the case, as the result of this transformation diverges in the limit \( T \rightarrow \infty \). 
Specifically, it is clear that in the limit \( T \rightarrow \infty \) the  correct imaginary-time propagator 
$
\langle \tilde{\bm{x}} | e^{-\hat{H}/(k_B T)} | \bm{x} \rangle
$ becomes $\delta(\tilde{\bm{x}}-\bm{x})$, and thus is zero for $\tilde{\bm{x}}\neq\bm{x}$, but the result given by the \( t \rightarrow -i\tau \)  transformation diverges in this limit,  even for $\tilde{\bm{x}}\neq\bm{x}$. In Appendix~\ref{pro}, we demonstrate this again with detailed calculations. This problem is crucial, as the semi-classical approximation should be applicable in the high-temperature limit. 
Although several alternative IVRs for the imaginary-time propagator have been derived \cite{Zhao and Miller, Yan and Shao}, an IVR based on coherent-state-type wave functions, similar to the HK IVR, has yet to be found.
 
 In this work, we solve the above problem for systems 
 and derive a reasonable HK-like semi-classical IVR for the Boltzmann operator,
 via an approach generalized from the one of Kay in Ref.~\cite{Kay}. Our result is applicable to systems with potential energy functions \( V(\mathbf{q}) \) of either of the following two types, where \( \mathbf{q} = (q_1, \dots, q_N) \) denotes the coordinates:

 \begin{itemize}
 
\item[] {\bf Type (a)}:  The norm of the potential energy gradient (i.e., the force intensity), \( |\nabla_{\bm q}V({\bm q})| \), has a finite upper bound.
 
\item[] {\bf Type (b)}:  The potential behaves harmonically in the long-range limit. That is, 
the potential can be expressed as \(  V({\bm q}) = \tilde V({\bm q}) + \sum_{j=1}^N \alpha_j q_j^2 \), with $\alpha_j$ being constants, and the norm of the gradient of  \( \tilde V({\bm q}) \) also has a finite upper bound.

\end{itemize}

\begin{figure}[tbp]
			\includegraphics[width=1.05\columnwidth]{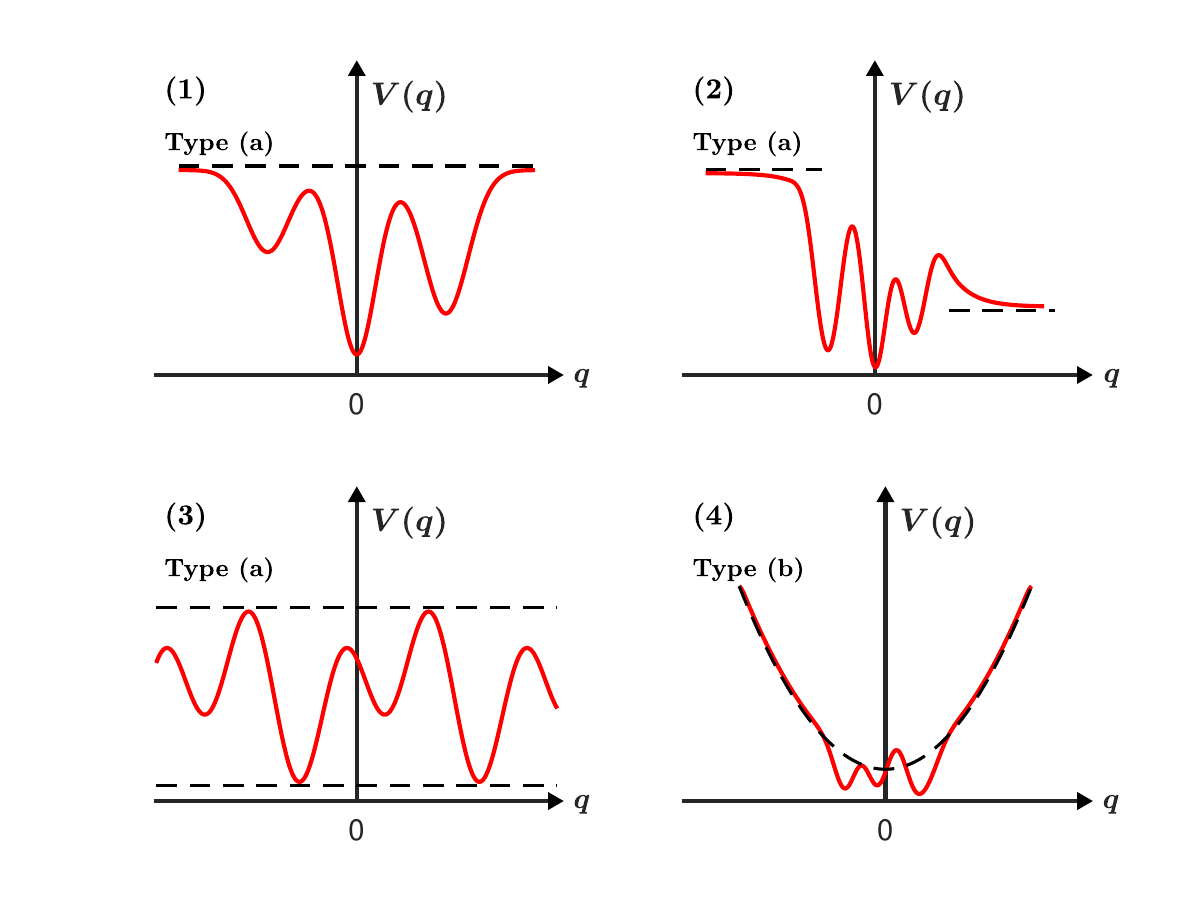}
			\caption{Schematic diagrams of some typical potentials (red solid lines)  of types (a) or (b) of Sec.~\ref{introduction}, for a one-dimensional system with coordinate $q$.  All these potentials are smooth functions of $q$. {\bf (1-3):} Typical potentials of type (a). Specifically, the potential in (1)  approaches to the same constant (dashed line) in the limits $q\rightarrow +\infty$ and $-\infty$, the potential in (2) approaches to the different constants (dashed lines) in these two limits, and the potential in (3) exhibits oscillatory behavior over the  entire $q$-axis, with finite upper and lower bounds (dashed lines).
		{\bf (4):} A typical potential of type (b), which approaches to a harmonic potential (dashed line) in the limits $q\rightarrow\pm\infty$.}
			\label{fig1}
		\end{figure}

 \noindent Fig.~\ref{fig1} illustrates schematic diagrams of several typical potentials of types (a) or (b). For systems with potential of these types,  in the high-temperature limit our IVR can produce the correct imaginary-time propagator 
 $\delta(\tilde{\bm x}-\bm x)$. Furthermore, our IVR is exact for free particles and harmonic oscillators.
The applicability for other systems are also illustrated via numerical examples.  
Our results are helpful for studies of the properties of atomic systems, molecular systems, and chemical reactions, at finite temperatures.

 Given the importance of studies on thermal properties of quantum systems, many researchers have developed various efficient approaches for calculating the imaginary-time propagator and other related physical quantities, such as the time-evolving Gaussian approximation \cite{mandelshtam01,mandelshtam02,Pollak02}, coherent state or Gaussian series representation methods \cite{Pollak03,Pollak04,Pollak05,Pollak06}, and polynomial expansion or fast Fourier transform methods \cite{metiu02,metiu03,metiu01}. 
Our approach can contribute to studies of finite-temperature quantum effects together with these existing methods.

The remainder of this paper is organized as follows. For the convenience of the readers, we first display our HK-like IVR for the Boltzmann operator  in Sec.~\ref{res}, 
and then show the derivation of this IVR as well as the condition for the semi-classical approximation in  Sec.~\ref{de}.
The applicability of our HK-like IVR is illustrated with some examples in Sec.~\ref{exa}.
 In Sec.~\ref{sum} there is a summary. Some details of the derivations and calculations are given in the appendixes.

\section{Central Result}
\label{res}

We consider a general multi-particle quantum system, with 
Cartesian components of the coordinates and momenta being denoted as 
$({p}_{1},...,{p}_{N})$ and $({q}_{1},...,{q}_{N})$, respectively, and the Hamiltonian being given by 
\begin{equation}
H=\sum_{j=1}^{N}\frac{{p}_{j}^{2}}{2m_{j}}+V({\bm q}).\label{h}
\end{equation}
Here
${\bm q}=({q}_{1},...,{q}_{N})$, and
 $m_{j}$ ($j=1,...,N$) is mass of the particle to which the coordinate $q_{j}$ belongs. Furthermore,  the potential $V({\bm q})$ is of either type (a) or type (b) of Sec.~\ref{introduction}. 
 
The Boltzmann
operator of our system is $e^{-\hat{H}/(k_{B}T)}$, where $\hat{H}$ is the quantum operator of the  Hamiltonian in Eq.~(\ref{h}).
The
matrix element
of this operator in the coordinate representation 
 (imaginary-time propagator) can be expressed as
\begin{eqnarray}
K_{\tau}(\tilde{\bm {x}},{\bm x}):=
\langle\tilde{\bm{x}}|e^{-\hat{H}/(k_{B}T)}|\bm{x}\rangle=
\langle\tilde{\bm{x}}|e^{-\hat{H}\tau/\hbar}|\bm{x}\rangle,
\end{eqnarray}
where $|\bm{x}\rangle$ and $|\tilde{\bm{x}}\rangle$ are eigen-states
of the coordinate operators, with eigen-values $\bm{x}=(x_{1},...,x_{N})$
and $\tilde{\bm{x}}=(\tilde{x}_{1},...,\tilde{x}_{N})$, respectively, and 
\begin{align}
\tau=\frac{\hbar}{k_{B}T}.\label{tau}
\end{align}

The HK-like IVR for the Boltzmann operator under the semi-classical approximation, which we have derived in this work, can be expressed as:
\begin{eqnarray}
K_{\tau}(\tilde{\bm {x}},{\bm x}) & = &A\int d\bm{p}d\bm{q}\left[D_\tau e^{-\frac{1}{\hbar}(S_\tau+B_\tau+C_\tau)}\right].
\label{asu2}
\end{eqnarray}
Here, ${\bm q}=({q}_{1},...,{q}_{N})$ (as defined above), $\bm{p}=(p_1,...,p_N)$, and  $\int d\bm{p},d\bm{q}=\int dp_{1}\cdots dp_{N},dq_{1}\cdots dq_{N}$. Moreover,
the factors $S_\tau$, $A$, $B_\tau$, $C_\tau$ and $D_\tau$ are all {\it real} and independent of $\hbar$, with the definitions being introduced in the following.

\subsection*{Factor $S_\tau$}

The factor $S_\tau$ of Eq.~(\ref{asu2}) is a  function of $\bm{p}$ and $\bm{q}$, and is independent of ${\bm x}$ and $\tilde{\bm x}$. It
is defined as
\begin{equation}
S_\tau(\bm{q},\bm{p})=\int_{0}^{\tau}d\eta\bigg[\sum_{j=1}^{N}\frac{{{p}}_{\eta,j}^{2}}{2m_{j}}+V({q}_{\eta,1},...,{q}_{\eta,N})\bigg].\label{s}
\end{equation}
Here  $q_{\eta,1},...,q_{\eta,N}$ and $p_{\eta,1},...,p_{\eta,N}$
satisfy the Hamilton's equations with inverted potential $-V$, initial position  ${\bm q}$ and initial momenta ${\bm p}$, i.e., the equations
\begin{align}
\frac{d}{d\eta}{q}_{\eta,j} & =\frac{{p}_{\eta,j}}{m_{j}};\label{h1}\\
\frac{d}{d\eta}{p}_{\eta,j} & =\frac{\partial V(z_{1},...,z_{N})}{\partial z_{j}}
\bigg\vert_{z_1=q_{\eta, 1};...;z_N=q_{\eta, N}},\label{h2}\\
&(j=1,...,N),\nonumber 
\end{align}
and the initial conditions 
\begin{align}
q_{\eta=0,j}=q_{j};\ \ \ p_{\eta=0,j}={p_{j}}, \hspace{0.5cm} (j=1,...,N). \label{ice}
\end{align}

According to the above definitions and equations, the factor $S_\tau$ of Eq.~(\ref{s})
is the action of the classical trajectory $q_{\eta,1},...,q_{\eta,N}$ and $p_{\eta,1},...,p_{\eta,N}$.
Moreover, in the following
we will consider $q_{\eta,1},...,q_{\eta,N}$ and $p_{\eta,1},...,p_{\eta,N}$ as functions
of $\eta$ and $\{{\bm q},{\bm p}\}$.

\subsection*{Factors $A$,  $B_\tau$ and $C_\tau$}

The factors $B_\tau$ and $C_\tau$ of Eq.~(\ref{asu2}) are  functions of both ${\bm p},\bm q$ and $\bm x, \tilde{\bm x}$. They are defined as
\begin{align}
B_\tau ({\bm p},\bm q;\bm x, \tilde{\bm x}) & =\sum_{j=1}^{N}\bigg[\gamma_j\left({x}_{j}-{q}_{j}\right)^{2}+\gamma_j\left(\tilde{{x}}_{j}-{{q}}_{\tau,j}\right)^{2}\bigg.\nonumber\\
 & \ \ \ \ \ \ \ \ \ \ \bigg.-{p}_{j}\left({x}_{j}-{q}_{j}\right)+{{p}}_{\tau,j}\left(\tilde{{x}}_{j}-{{q}}_{\tau,j}\right)\bigg];\label{biga}\\
C_\tau ({\bm p},\bm q;\bm x, \tilde{\bm x})& =\frac{1}{\tau}\sum_{j=1}^{N}m_{j}\bigg[\left(\tilde{{x}}_{j}-{{q}}_{\tau,j}\right)-\left({x}_{j}-{q}_{j}\right)\bigg]^{2},\label{bigb}
\end{align}
where $\gamma_{1,...,N}$ are arbitrary positive parameters. Additionally, the factor $A$ is defined as:
\begin{eqnarray}
A=\prod_{j=1}^N\left(\frac{\gamma_j}{2\hbar^{3}\pi^{3}}\right)^{1/2}.\label{cast}
\end{eqnarray}
Notice that the factor $C_\tau$ cannot be obtained from direct 
$t\rightarrow -i\tau$ transformation on
the HK representation of real-time propagator.

\subsection*{Factor $D_\tau$}

Similar to  $S_\tau$, the factor $D_\tau$  in Eq.~(\ref{asu2}) is also a
 function of ${\bm p},\bm q$, and is independent of $\bm x$ and $\tilde{\bm x}$.
To show the definition of $D_\tau$, we first introduce four $N\times N$ matrices 
${\rm R}^{qu}(\eta)$, ${\rm R}^{qv}(\eta)$, ${\rm R}^{pu}(\eta)$, and ${\rm R}^{pv}(\eta)$, with elements:
\begin{eqnarray}
{\rm R}_{ij}^{qu}(\eta) & = & -\frac{2}{\eta}m_{i}\delta_{ij}+\left(2\gamma_j+\frac{2}{\eta}m_{j}\right)\frac{\partial q_{\eta,j}}{\partial q_{i}}-\frac{\partial p_{\eta,j}}{\partial q_{i}};\nonumber \\
\label{dr1}\\
{\rm R}_{ij}^{qv}(\eta) & = & 2\gamma_j\delta_{ij}-\frac{2}{\eta}m_{j}\left(\frac{\partial q_{\eta,j}}{\partial q_{i}}-\delta_{ij}\right);\\
{\rm R}_{ij}^{pu}(\eta) & = & \left(2\gamma_j+\frac{2}{\eta}m_{j}\right)\frac{\partial q_{\eta,j}}{\partial p_{i}}-\frac{\partial p_{\eta,j}}{\partial p_{i}};\\
{\rm R}_{ij}^{pv}(\eta) & = & \delta_{ij}-\frac{2}{\eta}m_{j}\frac{\partial q_{\eta,j}}{\partial p_{i}},\label{r4-1}\\
&&(i,j=1,...,N),
\end{eqnarray}
with $\delta_{ij}$ being the Kronecker symbol. We further define other four $N\times N$ matrices 
${\rm T}^{uq}(\eta)$, ${\rm T}^{vq}(\eta)$, ${\rm T}^{up}(\eta)$, and ${\rm T}^{vp}(\eta)$, which relate to the these R-matrices via
\begin{equation}
\left(
\begin{array}{cc}
{\rm T}^{uq}(\eta) & {\rm T}^{up}(\eta)\\
{\rm T}^{vq}(\eta) & {\rm T}^{vp}(\eta)
\end{array}
\right)
=
\left(
\begin{array}{cc}
{\rm R}^{qu}(\eta) & {\rm R}^{qv}(\eta)\\
{\rm R}^{pu}(\eta) & {\rm R}^{pv}(\eta)
\end{array}
\right)^{-1}.\label{dt}
\end{equation}

The factor $D_\tau$ of Eq.~(\ref{asu2}) can be expressed in terms of the  elements of the above T-matrices, which are denoted as
${\rm T}^{to}_{ij}(\eta)$ ($t=u,v$; $o=p,q$; $i,j=1,...,N$). Specifically, we have
\begin{align}
D_\tau(\bm{q},\bm{p})= e^{-\int_{0}^{\tau}g_{\eta}(\bm{q},\bm{p})d\eta},
\label{bigc}
\end{align}
with $g_{\eta}$ being a function of $(\bm{q},\bm{p})$:
\begin{align}
&g_{\eta}(\bm{q},\bm{p})\nonumber \\
= & -\frac{1}{2}\sum_{i,j=1}^{N}{\rm W}_{ij}(\eta)V_{ij}(\eta)\nonumber \\
 & +\sum_{j=1}^{N}\bigg\{\left(\frac{1}{\eta}+\frac{\gamma_j}{m_{j}}\right)+\left(\frac{2\gamma_j^{2}}{m_{j}}+\frac{m_{j}}{\eta^{2}}+\frac{4\gamma_j}{\eta}\right){\rm W}_{jj}(\eta)\nonumber \\
 & \ \ \ \ \ \ \ \ \ \ +\left(\frac{2m_{j}}{\eta^2}+\frac{4\gamma_j}{\eta}\right){\rm T}_{jj}^{uq}(\eta)-\frac{m_{j}}{\eta^2}{\rm T}_{jj}^{vq}(\eta)\bigg\},\nonumber \\
 \label{ftau}
\end{align}
where
\begin{align}
V_{ij}(\eta)&=\left.\frac{\partial^{2}}{\partial z_{i}\partial z_{j}}V(z_{1},...,z_{N})\right\vert_{z_1=q_{\eta,1};...;z_N=q_{\eta,N}},\\
{\rm W}_{ij}(\eta) & =-\sum_{s=1}^{N}\left[{\rm T}_{is}^{uq}(\eta)\frac{\partial q_{\eta,j}}{\partial q_{s}}+{\rm T}_{is}^{up}(\eta)\frac{\partial q_{\eta,j}}{\partial p_{s}}\right].
\end{align}
According to this definition, calculating \( D_\tau \) requires computing the derivatives 
 $\frac{\partial q_{\eta,i}}{\partial q_j}$, $\frac{\partial q_{\eta,i}}{\partial p_j}$, $\frac{\partial p_{\eta,i}}{\partial q_j}$ and $\frac{\partial p_{\eta,i}}{\partial p_j}$ ($i,j=1,...,N$)  for $0\leq\eta\leq \tau$. 
  These functions can be derived by solving Hamilton's equations (\ref{h1}, \ref{h2}) in conjunction with another set of ordinary differential equations, as demonstrated in, for example, Eqs.~(10a-10d) of Ref.~\cite{Kluk1986}.
 
  Note that the determinant of the matrix 
 \begin{eqnarray}
 \left(
\begin{array}{cc}
{\rm R}^{qu}(\eta) & {\rm R}^{qv}(\eta)\\
{\rm R}^{pu}(\eta) & {\rm R}^{pv}(\eta)
\end{array}
\right)\nonumber
\end{eqnarray}
 may becomes zero when ${\bm q}$, ${\bm p}$ and $\eta$ take certain specific values.
 As a result, the function $g_\eta({\bm q}, {\bm p})$, which depends on the inverse of this matrix, diverges for these specific ${\bm q}$, ${\bm p}$ and $\eta$.
If for specific ${\bm q}$ and ${\bm p}$ (denoted as ${\bm q}_\ast$ and ${\bm p}_\ast$, respectively) the function
$g_\eta({\bm q}_\ast, {\bm p}_\ast)$ diverges at $\eta=\eta_1,\eta_2,..,\eta_{n}$, ($\eta_{1,...,n}\in (0,\tau)$), then for
the corresponding factor $D_\tau(\bm{q}_\ast,\bm{p}_\ast)$,
 the 
 expression (\ref{bigc}) 
 should be replaced by
 \begin{eqnarray}
 D_\tau(\bm{q}_\ast,\bm{p}_\ast)= \cos\left(\pi \tilde g\right) \cdot
 e^{-{\cal P}\int_{0}^{\tau}g_{\eta}
 (\bm{q}_\ast,\bm{p}_\ast)
 d\eta},
\label{bigcp}
 \end{eqnarray}
 where $\cal P$ denotes the Cauchy principal value integral, and the factor $\tilde g$ is defined as 
 \begin{eqnarray}
\tilde g= \sum_{\xi=1}^{n}\lim_{\eta\rightarrow\eta_\xi}\bigg[(\eta-\eta_\xi) \cdot g_{\eta}
 (\bm{q}_\ast,\bm{p}_\ast)\bigg].
 \label{gtilde2}
 \end{eqnarray}
Clearly, if $g_{\eta}
 (\bm{q}_\ast,\bm{p}_\ast)$ did not diverge for any $\eta\in(0,\tau)$, we  have $\tilde g=0$, and thus Eq.~(\ref{bigcp}) naturally returns to Eq.~(\ref{bigc}).

Note that the pre-factor in the HK representation of the real-time propagator, which is the counter part of the factor $D_\tau$, can be expressed as either the exponential of a function integrated over time (similar to Eq.~(\ref{bigc}))  \cite{Kay}, or simply as a determinant  \cite{HK,Kay}.
However, we cannot express the factor $D_\tau$  as a determinant.
On the other hand, the numerical calculations of  $D_\tau$ and  that pre-factor both rely on  computing the evolution of coordinates and momentums, as well as the derivatives of them on their initial values. Consequently, the  cost 
of these numerical calculations have the same scaling with the amount $N$ of the degree-of-freedoms.



\section{Derivation of Eq.~(\ref{asu2})}
\label{de}

Now we show our approach for deriving the HK-like IVR for the imaginary-time propagator, i.e., Eq.~(\ref{asu2}).
For simplicity,  we consider the system of a single particle in one-dimensional space ($N=1$), and the derivation can be directly generalized to the cases with arbitrary $N$.
 Consequently, we will omit the subscript denoting the particle index in the following. In the following we present the main ideas and framework of the derivation. The details of the derivations are provided in Appendix~\ref{d1b}.
 
To derive Eq.~(\ref{asu2}), we express the imaginary-time propagator as the integration
 \begin{eqnarray}
 K_\tau(\tilde x,x)=A\int dpdq\left[ F_D e^{-\frac{(S_\tau+B_\tau+C_\tau)}{\hbar}}\right],
 \label{test}
 \end{eqnarray}
 with $A$, $S_\tau$, $B_\tau$ and $C_\tau$ being defined in Eqs. (\ref{cast}), (\ref{s}), (\ref{biga}) and (\ref{bigb}), respectively, and $F_D$ being a to-be-determined function of $\tau$ and $(p,q)$, which is independent of $x$ and $\tilde x$.

 
 In the following, we first prove that the function $F_D$ satisfies the ``initial condition" $\lim_{\tau\rightarrow 0}F_{D}=1$.
Then we derive the differential equation of $F_D$, and solve this equation  together with this ``initial condition",
 using the semi-classical approximation. We will find that the solution is just $F_D=D_\tau$, with $D_\tau$ being given by  Eq.~(\ref{bigc}) for $N=1$.
 Substituting this result  into Eq.~(\ref{test}), we  obtain the result of (\ref{asu2}) for $N=1$.

\subsection{ ``Initial Condition" of $F_{D}$}

To prove $\lim_{\tau\rightarrow 0}F_{D}=1$, we first calculate the 
 limitation
$\lim_{\tau\rightarrow 0}A\int dpdq\  e^{-\frac{(S_\tau+B_\tau+D_\tau)}{\hbar}}$.
As shown in Appendix \ref{pic}, when the potential $V(q)$ is of types (a) or (b) of Sec.~I, we have
\begin{eqnarray}
\lim_{\tau\rightarrow 0}A\int dpdq \  e^{-\frac{(S_\tau+B_\tau+C_\tau)}{\hbar}}&=&\delta(x-\tilde x).\label{int2}
\end{eqnarray}

On the other hand, it is clear that $\lim_{\tau\rightarrow 0} K_\tau(\tilde x,x)=\delta(x-\tilde x)$. Thus, we have 
\begin{eqnarray}
\lim_{\tau\rightarrow 0} K_\tau(\tilde x,x)=\lim_{\tau\rightarrow 0}A\int dpdq\ e^{-\frac{(S_\tau+B_\tau+C_\tau)}{\hbar}}.
\end{eqnarray}
Comparing this result and Eq.~(\ref{test}), we find that the function $F_D$ satisfies
 \begin{eqnarray}
 \lim_{\tau\rightarrow 0}F_{D}=1.\label{ic}
 \end{eqnarray}

\subsection{ Equation of $F_D$ and Semi-Classical Approximation}

Now we derive the differential equation for the function $F_D$. To this end, we introduce the correction operator \cite{co1,co2,co3}
 for our system, which is defined as
 \begin{eqnarray}
 \hat{\Lambda}_{\tilde{x}}:=
 \hbar\frac{\partial}{\partial\tau}-\frac{\hbar^{2}}{2m}\frac{\partial^{2}}{\partial\tilde{x}^{2}}+V(\tilde{x})
 \label{lam2}.
 \end{eqnarray}
It is clear that the imaginary-time propagator \( K_\tau(\tilde{x}, x) \) satisfies \( \hat{\Lambda} \left[ K_\tau(\tilde{x}, x) \right] = 0 \). By substituting Eq.~(\ref{test}) into this equation, we find that the function \( F_D \) satisfies
\begin{eqnarray}
 \hat{\Lambda}\bigg[\int dpdq\left( F_D e^{-\frac{(S_\tau+B_\tau+C_\tau)}{\hbar}}\right)\bigg]=0.\label{seqa1}
\end{eqnarray}
 As detailed in Appendix~\ref{proof}, using the method generalized from Ref.~\cite{Kay}, we find that the sufficient condition for Eq.~(\ref{seqa1}) can be expressed as a differential equation for \( F_D \):
 \begin{eqnarray}
\bigg(\hat {\tilde L}_0+\hbar \hat {\tilde L}_1+\hbar^2\hat {\tilde L}_2+...\bigg)F_D=0,\label{eqf}
 \end{eqnarray}
 where $\hat {\tilde L}_{0,1,2,...}$ are $\hbar$-independent operators. Specially, the operator $\hat {\tilde L}_0$ is given by
 \begin{eqnarray}
\hat {\tilde L}_0=\frac{\partial }{\partial \tau}+g_\tau(q,p),
 \end{eqnarray}
where $g_\tau(q,p)$ is just the function given by Eq.~(\ref{ftau}) with $N=1$.


 \begin{figure*}[tbp]
			\includegraphics[width=1.7\columnwidth]{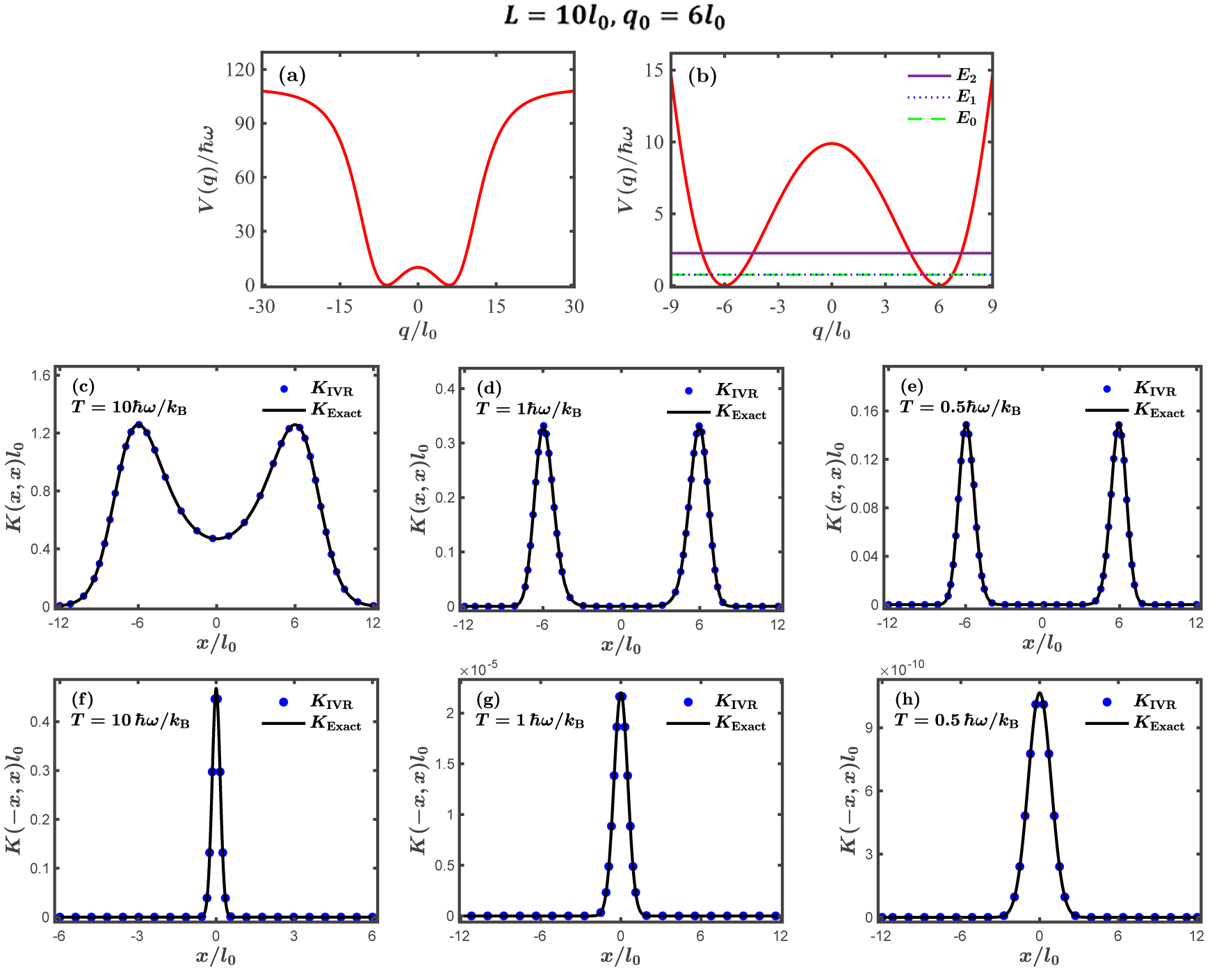}
			\caption{
			The potential $V(q)$ of Eq.~(\ref{vq}), and
			matrix elements of the Boltzmann operator, for  cases with $L=10l_0$ and $q_0=6l_0$.
			{\bf (a)}: The potential $V(q)$. {\bf (b)}: Enlarged view of the region around $q=0$ in (a). Here we also show the ground-state energy $E_0$ (green dashed line), the first excited-state energy $E_1$ (blue dotted line), and the second excited-state energy $E_2$ (purple solid line), which are given by numerical diagonalization  of the Hamiltonian. Note that $E_0$ and $E_1$ lie very close to each other.			{\bf (c-e)}: The diagonal elements $K(x,x)=\langle x|e^{-\hat{H}/(k_{B}T)}|x\rangle$ (in units of $1/l_0$). {\bf (f-h)}: The non-diagonal elements $K(-x,x)=\langle -x|e^{-\hat{H}/(k_{B}T)}|x\rangle$  (in units of $1/l_0$). Here we show the results with temperature $T=10\hbar\omega/k_B$ (c, f), $T=\hbar\omega/k_B$ (d, g) and $T=0.5\hbar\omega/k_B$ (e, h). For each temperature, we illustrate the results $K_{\rm Exac}$ from exact diagonlization of the Hamiltonian (black dotted line) and the results $K_{\rm IVR}$ given by our HK-like IVR (blue dots).
			}
			\label{x4}
		\end{figure*}

Note that the operator in the l.h.s of Eq.~(\ref{eqf}) is expanded as a power series of $\hbar$.
Under the semi-classical approximation, we further ignore the terms proportional to $\hbar^{n}$ ($n \geq 1$) in this series, and approximate Eq.~(\ref{eqf}) as
\begin{eqnarray}
\bigg[\frac{\partial }{\partial \tau}+g_\tau(q,p) \bigg]F_D(q,p,\tau)=0.\label{eqf2}
\end{eqnarray}
Furthermore, it is clear that under the initial condition (\ref{ic}), the solution of Eq.~(\ref{eqf2}) is just
 \begin{eqnarray}
F_D(q,p,\tau)= e^{-\int_{0}^{\tau}g_{\eta}(q,p)d\eta}.\label{fcr}
 \end{eqnarray}
  Additionally, at the end of Appendix~\ref{fddiv}, we further prove that 
 if for specific $q$, $p$ (denoted as $q_\ast$, $p_\ast$),
 $g_{\eta}(q_\ast,p_\ast)$
diverges when $\eta=\eta_1,\eta_2,...,\eta_n$ ($\eta_{1,...,{n}}\in (0,\tau)$), then for $F_D(q_\ast,p_\ast,\tau)$, the expression (\ref{fcr}) should be replaced by  
 \begin{eqnarray}
 F_D(q_\ast,p_\ast,\tau)= \cos\left(\pi \tilde g\right) \cdot
 e^{-{\cal P}\int_{0}^{\tau}g_{\eta}
 (q_\ast,p_\ast)
 d\eta},
\label{bigcp2}
 \end{eqnarray}
 with 
 \begin{eqnarray}
\tilde g= \sum_{\xi=1}^{n}\lim_{\eta\rightarrow\eta_\xi}\bigg[(\eta-\eta_\xi) \cdot g_{\eta}
 (q_\ast,p_\ast)\bigg].
 \label{gtilde22}
 \end{eqnarray}

 
 \subsection{Final Derivation}

Eq.~(\ref{fcr}) (or Eq.~(\ref{bigcp2})) yields that 
 $
F_D=D_\tau,
$
 with $D_\tau$ being the one defined in Eq.~(\ref{bigc}) (or Eq.~(\ref{bigcp})) with $N=1$. 
Substituting this result into Eq.~(\ref{test}), we finally obtain  Eq.~(\ref{asu2}) for $N=1$.

 \subsection{Condition of the Semi-Classical Approximation}
 \label{condition}
 
 \begin{figure*}[tbp]
			\includegraphics[width=1.7\columnwidth]{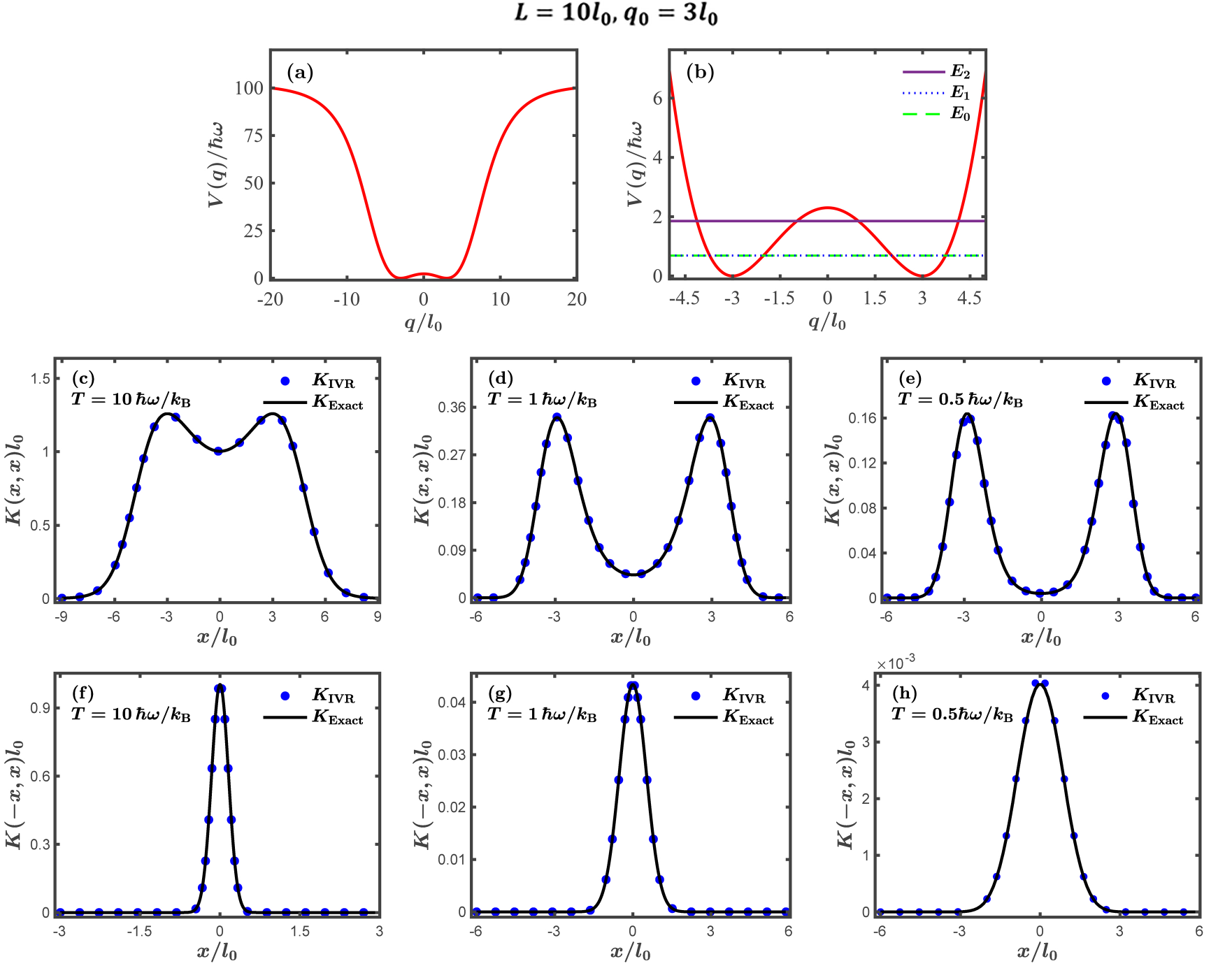}
			\caption{
			Same as Fig.~(\ref{x4}), but for  $L=10l_0$ and $q_0=3l_0$.
			}
			\label{x8}
		\end{figure*}
 
At the end of this section, we discuss the condition underlying the semi-classical approximation employed in deriving our HK-like IVR, namely, the condition under which our HK-like IVR provides a good approximation to the exact matrix element of the Boltzmann operator. For comparison, recall that the semi-classical approximation used in the derivation of the van Vleck propagator requires the characteristic length (CL) scale of the potential energy variation to be much larger than the de Broglie wavelength~\cite{Weinberg}. The condition adopted here is similar: 

\begin{eqnarray}
\text{classical\ CLs}\gg\text{quantum\ CLs}.\label{con}
\end{eqnarray}
Here classical CLs refer to those associated with the potential energy, which are
$\hbar$-independent, and quantum CLs are those proportional to $\hbar^\alpha$ with $\alpha>0$. 
Specifically, for a system with $N$ coordinates, the quantum CLs are $\sqrt{\hbar/\gamma_j}$ and $\sqrt{\hbar\tau/m_j}=\sqrt{\hbar^2/(k_BTm_j)}$ ($j=1,...,N$), etc. Note that condition (\ref{con}) implies that the semi-classical approximation is not applicable at very low temperatures. Additionally,
in practical calculations for realistic systems,
appropriate values of 
 $\gamma_j$ ($j=1,...,N$)
should be chosen  to ensure that the condition Eq.~(\ref{con}) is satisfied.

\section{Examples}
\label{exa}

In the previous two sections, we presented our HK-like IVR for the imaginary-time propagator and its derivation. In this section, we demonstrate its applicability. Specifically, we analytically prove that the IVR is exact for free particles and harmonic oscillators, and numerically illustrate its performance with an anharmonic system.

\subsection{Free Particles}

We first consider the system of $N$ free particles, i.e., $V=0$. As mentioned above, in this case the solutions of the Hamilton's equations (\ref{h1}, \ref{h2}) are just $q_{\eta,j}=q_j+\eta p_j/m_j$ and $p_{\eta,j}=p_j$ ($j=1,...,N$). Consequently, the integration in the r. h. s. of Eq.~(\ref{asu2}) can be performed analytically.
With straightforward calculations, we find that 
  the right-hand side (r. h. s.) of Eq.~(\ref{asu2}) is 
  \begin{eqnarray}
\prod _{j=1}^N 
\left({\frac{m_j}{2\pi\hbar\tau}}\right)^{1/2}
e^{-\frac{m_j}{2\hbar\tau}\left(x_j-\tilde x_j\right)^2},
\end{eqnarray}
which is same as the  exact imaginary-time propagator  for this system. Thus, our HK-like IVR 
is exact for the free particles.

\subsection{Harmonic Oscillators}

Now we consider $N$ harmonic oscillators with potential energy $V=\sum_{j=1}^N m_j\omega_j^2 x_j^2/2$. The Hamilton's equations (\ref{h1}, \ref{h2}) for this system can also be solved analytically, and we have $q_{\eta,j}=c_je^{\omega_j\eta}+d_je^{-\omega_j\eta}$ and $p_{\eta,j}=m_j\omega_j(c_je^{\omega\eta}-d_je^{-\omega\eta})$, where $c_j=(p_j +m_j \omega_j q_j)/(2m_j\omega_j)$ and $d_j=(m_j \omega_j q_j-p_j)/(2m_j\omega_j)$. Using these results, we can also analytically perform the integration in the r. h. s. of Eq.~(\ref{asu2}), and find that the r. h. s. of 
Eq.~(\ref{asu2}) is just (Appendix~\ref{ho})
\begin{eqnarray}
\prod _{j=1}^N\left(\frac{m_j\omega_j}{2\pi\hbar\sinh(\omega_j \tau)}\right)^{1/2}e^{-\frac{m_j\omega_j\left[
\left(x_j^2+\tilde x_j^2\right)\cosh(\omega_j \tau)-2x_j\tilde x_j
\right]}{2\hbar\sinh(\omega_j \tau)}},\nonumber\\
\label{ho3}
\end{eqnarray}
which is same as the  exact imaginary-time propagator  of these oscillators.
Therefore, as for the free particles, our IVR is also exact for harmonic oscillators.

\subsection{Anharmonic System }

Finally, we consider a single particle in an anharmonic potential,  
with Hamiltonian $H = p^2/(2m) + V(q)$. Here $m$, $p$ and $q$ are the mass, momentum and coordinate of this particle, respectively. The anharmonic potential $V(q)$ is given by
\begin{equation}
V(q)=-\frac{m\omega^2L^2}{1-\frac{q^2}{2L^2}+
\frac{q^4}{4L^2q_0^2}}+V_0,\label{vq}
\end{equation}
where $\omega$, $q_0$, and $L$ are positive parameters, and $V_0$ is a $q$-independent constant:
\begin{equation}
V_0=\frac{m\omega^2 L^2}{1-\frac{q_0^2}{4L^2}},\label{v0}
\end{equation}
which is chosen such that the minimum value of $V(q)$ is zero.
Specifically, 
$\omega$ has the dimension of frequency, and both $q_0$ and $L$ have the dimension of length, and satisfy $q_0<2L$. 
Clearly,  this potential is of type (a) of Sec.~I. Furthermore,
as shown in 
Fig.~\ref{x4}(a, b) and Fig.~\ref{x8}(a, b), for typical parameters,
in the region around $q=0$, $V(q)$ is a double-well  with minimum points being localized at $q=\pm q_0$.

Furthermore, note that according to Eq.~(\ref{con}), the semi-classical approximation is applicable when the classical CLs are much larger than the quantum CLs. For our system the classical CLs are $L$ and $q_0$, and the 
quantum CLs are:
\begin{eqnarray}
l_0=\sqrt{\hbar/(m\omega)};\ \ l_\gamma=\sqrt{{\hbar}/{\gamma}};\ \ l_T=\frac{\hbar}{\sqrt{k_BTm}}.\label{QCL}
\end{eqnarray}
Thus, the condition (\ref{con}) of the semi-classical approximation can be expressed as
\begin{eqnarray}
L, q_0\gg l_0,l_\gamma,l_T.\label{con2}
\end{eqnarray}
Due to this condition, we consider the systems with
\begin{itemize}
\item[{(i)}]: $L=10l_0$, $q_0=6l_0$;
\item[{(ii)}]: $L=10l_0$, $q_0=3l_0$, 
\end{itemize} 
and  temperatures  $T = {10\hbar \omega}/{k_B}, T={\hbar \omega}/{k_B}$ and $T={0.5\hbar \omega}/{k_B}$ (i.e., $l_T=0.32l_0, l_T=l_0,$ and $l_T=1.41 l_0$, respectively),
and choose the parameter $\gamma$ in our IVR as 
$\gamma =5 {\hbar}/{l_0^2}$ (i.e., $l_\gamma= 0.45l_0$).

For  systems with above parameters we calculate the matrix elements of the Boltzmann operator for this particle,
$K(\tilde{x}, x) \equiv \langle \tilde{x} | e^{-\hat{H}/(k_B T)} | x \rangle$,
via both our HK-like semi-classical  IVR and  exact numerical diagonalization of the Hamiltonian operator \(\hat{H}\). 
In Figs.~\ref{x4}(c-h) and \ref{x8}(c-h), we compare the \(K(\tilde{x}, x)\) obtained via
our IVR and exact diagonalization   approach, which are denoted as \(K_{\rm IVR}(\tilde{x}, x)\) and \(K_{\rm Exact}(\tilde{x}, x)\), respectively,
for \(\tilde{x} = \pm x\). 
It is shown that they are in perfect quantitative agreement.

 \begin{figure*}[tbp]
			\includegraphics[width=2.1\columnwidth]{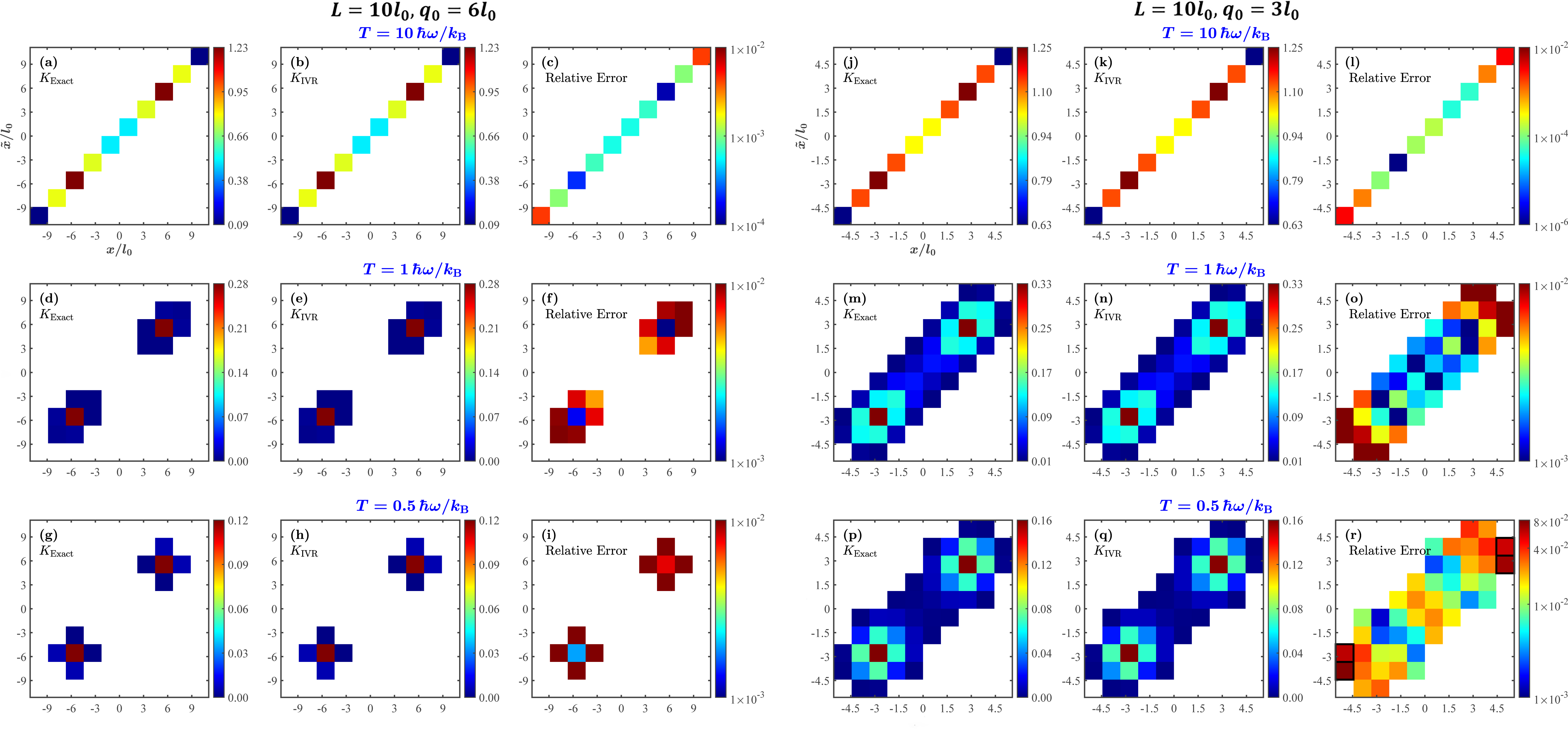}
			\caption{
			 $K_{\rm Exac}(\tilde x, x)$  (in units of $1/l_0$),  $K_{\rm IVR}(\tilde x, x)$  (in units of $1/l_0$), and 
		the	relative error of our HK-like IVR approach which is defined as 
${|K_{\rm Exact}(\tilde{x}, x) - K_{\rm IVR}(\tilde{x}, x)|}/{|K_{\rm Exact}(\tilde{x}, x)|}$.
Here we show the results for cases with  $L=10l_0$, $q_0=6l_0$ (a-i) and  $L=10l_0$, $q_0=3l_0$  (j-r), for temperatures $T=10\hbar\omega/k_B$, $T=\hbar\omega/k_B$, and $T=0.5\hbar\omega/k_B$. 
We do not show the results in the white regions, since in these regions both $K_{\rm Exact}$ and $K_{\rm IVR}$ are below 1\% of the maximum value of $K_{\rm Exact}$ across the entire domain (denoted as $K^{\rm max}_{\rm Exact}$). Moreover, in the four squares enclosed by black lines in (r), the relative error is between \( 4 \times 10^{-2} \) and \( 8 \times 10^{-2} \), while both $K_{\rm Exact}$ and $K_{\rm IVR}$ are below 3\% of $K^{\rm max}_{\rm Exact}$.	}
			\label{error}
		\end{figure*}

In Fig.~\ref{error}, we further compare \( K_{\rm IVR}(\tilde{x}, x) \) with \( K_{\rm Exact}(\tilde{x}, x) \) for additional values of \( \tilde{x} \) and \( x \), and present the relative error of our HK-like IVR approach, defined as 
$
{|K_{\rm Exact}(\tilde{x}, x) - K_{\rm IVR}(\tilde{x}, x)|}/{|K_{\rm Exact}(\tilde{x}, x)|}.
$
It is shown that for \( T = 10\hbar\omega / k_B \) and \( T = \hbar\omega / k_B \), the relative error remains below \( 10^{-2} \) for both parameters {(i)} and {(ii)}. Moreover, for \( T = 0.5\hbar\omega / k_B \), the relative error stays below \( 10^{-2} \) for parameter {(i)}, while it can reach up to \( 8 \times 10^{-2} \) for parameter {(ii)}. Specifically, for parameter {(ii)} with \( T = 0.5\hbar\omega / k_B \), the relative error is below \( 4 \times 10^{-2} \) in most of the \( (\tilde{x}, x) \) region, except at some places (the squares enclosed by black lines in Fig.~\ref{error}(r)) where \( K_{\rm Exact}(\tilde{x}, x) \) is very small compared to its maximum value in the entire \( (\tilde{x}, x) \) domain.
 Above results  demonstrate the applicability of our HK-like IVR method. 
 
 \begin{figure}[t]
			\includegraphics[width=0.6\columnwidth]{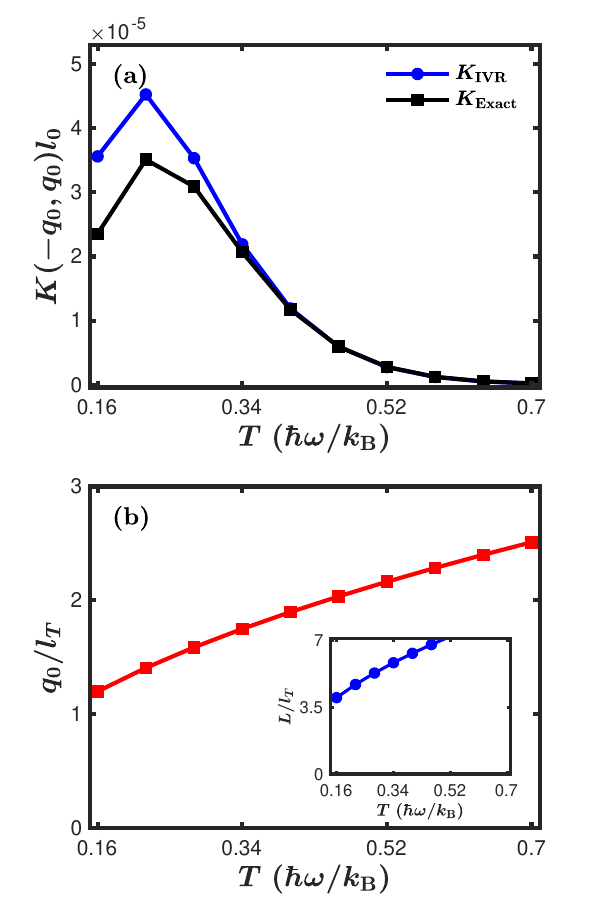}
			\caption{ \(K_{\rm IVR}(-q_0, q_0)\) and \(K_{\rm Exact}(-q_0, q_0)\) {\bf (a)},  
			and the ratio $q_0/l_T$ {\bf (b)},
			as functions of temperature $T$, for systems with parameter (ii). Inset of (b): the ratio $L/l_T$.}
			\label{addfig}
		\end{figure}
 
Furthermore, 
the relatively large relative error for parameter {(ii)} with 
$T = 0.5\hbar\omega / k_B$ can be understood as follows.
Eq.~(\ref{con2}) implies that the semi-classical approximation is applicable when $q_0\gg l_T$. However, 
 for parameter {(ii)} with 
$T = 0.5\hbar\omega / k_B$ we have $q_0=3l_0$ and $l_T=1.41l_0$, and thus this condition is not satisfied very well.

More generally, due to the condition \( L, q_0 \gg l_T \) shown in Eq.~(\ref{con2}), the semi-classical approximation becomes less accurate at very low temperatures. We illustrate this in Fig.~\ref{addfig}, where \( K_{\rm IVR}(-q_0, q_0) \) and \( K_{\rm Exact}(-q_0, q_0) \) are plotted as functions of temperature \( T \) in Fig.~\ref{addfig}(a) for systems with parameter set (ii). The corresponding values of \( q_0/l_T \) and \( L/l_T \) are shown in Fig.~\ref{addfig}(b). It is evident that, although the two quantities remain qualitatively consistent throughout, a significant quantitative difference emerges when \( T \lesssim 0.3\hbar\omega/k_B \) (\( q_0 \lesssim 1.5l_T \)), and this difference increases as \( T \) decreases. Note that in these low-temperature cases, methods based on Gaussian series representation \cite{mandelshtam01,mandelshtam02,Pollak04} can yield results that are more accurate than those obtained using our approach.

\section{Summary}
\label{sum}

In this work, we derive an HK-like semi-classical IVR (Eq.~(\ref{asu2})) for the Boltzmann operator, applicable to systems 
 with potentials of types (a) or (b) of Sec.~I.
The issue of non-convergence in the high-temperature limit present in the direct extension HK representation is overcome by our IVR.
Our IVR is exact for free particles and harmonic oscillators, and its applicability to other systems is demonstrated through examples.

 In addition to the potentials of types (a) or (b), our IVR may be also  applicable for systems with potentials 
with form \(  V({\bm q}) = \tilde V({\bm q}) + \sum_{j=1}^N \alpha_j q_j^{\nu_j} \), 
where $0<\nu_j\leq 2$ ($j=1,...,N$),
and $\alpha_j$ can be dependent on the sign of $q_j$, with the norm of the gradient of  \( \tilde V({\bm q}) \) having a finite upper bound.
For potentials with this behavior, but $\nu_j>2$ for some $j$, it is likely that our IVR is not applicable, as the integral in Eq.~(\ref{asu2}) may diverge, regardless of whether 
$\tau$ is zero or non-zero.

Our HK-like IVR can be used to calculate the partition function and various finite-temperature properties of molecular systems. It is worth noting that nearly all of these realistic systems are multi-dimensional (\(N > 1\)). For such systems, the majority of the numerical cost arises from the calculation of the pre-factor \(D_\tau\) in Eq.~(\ref{asu2}). Specifically, as shown above, deriving \(D_\tau\) requires calculating \(4N^2\) quantities, namely, \(\frac{\partial q_{\eta,i}}{\partial q_j}\), \(\frac{\partial q_{\eta,i}}{\partial p_j}\), \(\frac{\partial p_{\eta,i}}{\partial q_j}\), and \(\frac{\partial p_{\eta,i}}{\partial p_j}\) (\(i,j = 1,...,N\)). Additionally, note that other approaches for calculating the imaginary-time propagator, such as those based on Gaussian approximation \cite{mandelshtam01,mandelshtam02,Pollak02}, require the calculation of \(N^2\) quantities, rather than $4N^2$. Therefore, the cost of calculating \(D_\tau\) may offset the advantages of our method. In future work, we will explore approaches to reduce this cost further and apply our HK-like IVR to multi-dimensional realistic systems.



\section*{Acknowledgments}

We express our deep gratitude to Prof. Jiushu Shao for the insightful discussions and for providing numerous important suggestions.
This work is supported by the Innovation Program for Quantum Science and Technology (Grant No.~2023ZD0300700) and the National Key Research and Development Program of China (Grant No.~2022YFA1405300). 

\appendix

\onecolumngrid

\section{Divergence of Direct Extensions of HK Representation for $T\rightarrow\infty$.}
\label{pro}

In this appendix, we demonstrate that the imaginary-time propagator
$
\langle \tilde{\bm{x}} | e^{-\hat{H}/(k_B T)} | \bm{x} \rangle
$
obtained by applying the \( t \rightarrow -i\tau \) transformation to the HK representation~\cite{HK}, diverges in the high-temperature limit (\( T \rightarrow \infty \)), even for \( \tilde{\bm{x}} \neq \bm{x} \). Without loss of generality, we consider a system of a single particle in one-dimensional (1D) space. Specifically, we first investigate the general case with an arbitrary potential, and then illustrate the result for two specific examples: a free particle and a harmonic oscillator. Our analysis can be easily extended to systems with arbitrary dimensions and particle numbers.

\subsection{General Case}

We consider a 1D particle with mass $m$ and Hamiltonian $\hat H=\hat p^2/(2m)+V(\hat q)$, where $\hat p$ and $\hat q$ are momentum and coordinate operators, respectively.
The HK representation of the real-time propagator 
$\langle \tilde x|e^{-i\hat Ht/\hbar}|x\rangle$
of this particle is 
\begin{eqnarray}
\int dpdq 
\Theta(q,p,t)
e^{-\frac{\gamma(x-q)^2}{\hbar}-i\frac{p(x-q)}{\hbar}}e^{-\frac{\gamma(\tilde x-q_t)^2}{\hbar}+i\frac{p_t(\tilde x-q_t)}{\hbar}}e^{-i\frac{1}{\hbar}\int_0^t dt' \left[\frac{p_{t^\prime}^2}{2m}-V(q_{t^\prime})\right]},\label{HKf}
\end{eqnarray}
with
\begin{eqnarray}
\Theta(q,p,t)=\frac{1}{\sqrt \gamma}\left(
\frac \gamma2\frac{\partial p_t}{\partial p}
-i\gamma^2\frac{\partial q_t}{\partial p}
+\frac i4\frac{\partial p_t}{\partial q}
+\frac \gamma 2\frac{\partial q_t}{\partial q}
\right)^{1/2},
\end{eqnarray}
where $\gamma$ is an arbitrary positive number, and $q_{t}$ and $p_{t}$ satisfy the classical Hamiltonian equations 
\begin{eqnarray}
\frac{dq_{t}}{d{t}}=\frac {p_t}{m},\hspace{0.5cm}\frac {dp_{t}}{d{t}}=-\frac{\partial V(\eta)}{\partial \eta}\bigg|_{\eta=q_t},\label{hef}
\end{eqnarray}
with initial conditions $q_{t=0}=q$ and $p_{t=0}=p$.
Now we apply the transformation $t\rightarrow -i\tau$ to Eq.~(\ref{HKf}). Due to Eq.~(\ref{hef}), this transformation   leads to another transformations $q_{t}\rightarrow q_{\tau}$ and $p_{t}\rightarrow ip_{\tau}$, with $q_{\tau}$ and $p_{\tau}$ satisfying  
 \begin{eqnarray}
\frac{dq_{t}}{d{\tau}}=\frac {p_\tau}{m},\hspace{0.5cm}\frac {dp_{\tau}}{d{\tau}}=+\frac{\partial V(\eta)}{\partial \eta}\bigg|_{\eta=q_\tau},\label{hef2}
\end{eqnarray}
Substituting these results into Eq.~(\ref{HKf}), we find that 
the result of the $t\rightarrow -i\tau$ transformation  is  
\begin{eqnarray}
R(\tilde x,x,\tau)\equiv \int dpdq R_1(q,p,\tau)R_2(\tilde x, x,q,p,\tau),\label{bigr}
\end{eqnarray}
where 
\begin{eqnarray}
R_1(q,p,\tau)&=&\frac{1}{\sqrt \gamma}\left(
\frac \gamma2\frac{\partial p_\tau}{\partial p}
-\gamma^2\frac{\partial q_\tau}{\partial p}
-\frac 14\frac{\partial p_\tau}{\partial q}
+\frac \gamma 2\frac{\partial q_\tau}{\partial q}
\right)^{1/2},\label{bigr1}\\[8pt]
R_2(\tilde x, x,q,p,\tau)&=&e^{-\frac{\gamma(x-q)^2}{\hbar}+\frac{p(x-q)}{\hbar}}e^{-\frac{\gamma(\tilde x-q_\tau)^2}{\hbar}-\frac{p_\tau(\tilde x-q_\tau)}{\hbar}}  e^{-\frac{1}{\hbar}\int_0^\tau d\tau' \left[\frac{p_{\tau^\prime}^2}{2m}+V(q_{\tau^\prime})\right]}.
\label{bigr2}
\end{eqnarray}

Now we consider the behavior of $R(\tilde x,x,\tau)$ in the high-temperature limit 
$T\rightarrow \infty$ (i.e.,
$\tau\rightarrow 0$), for $\tilde x\neq x$. Due to the facts $R_1(q,p,\tau=0)=1$ and $R_2(\tilde x, x,q,p,\tau=0)=e^{-\frac{\gamma(x-q)^2}{\hbar}-\frac{\gamma(\tilde x-q)^2}{\hbar}+\frac{p(x-\tilde x)}{\hbar}}$, we have
\begin{eqnarray}
R(\tilde x,x,\tau\rightarrow 0)=\int dpdq e^{-\frac{\gamma(x-q)^2}{\hbar}-\frac{\gamma(\tilde x-q)^2}{\hbar}+\frac{p(x-\tilde x)}{\hbar}}=\infty,\hspace{0.5cm}{\rm even\ for\ }\tilde x\neq x.
\end{eqnarray}
I.e., the result of the $t\rightarrow -i\tau$ transformation to the HK representation, diverges in the high-temperature limit.

Additionally, we note that in Ref.~\cite{Miller}, the authors propose two other  imaginary-time extensions of the HK representation, which are also based on the \( t \rightarrow -i\tau \) transformation. These extensions  have similar forms to those in Eq.~(\ref{bigr}-\ref{bigr2}), but the signs of the terms \( \frac{p(x - q)}{\hbar} \) and \( \frac{p_\tau(\tilde{x} - q_\tau)}{\hbar} \) in the exponent of Eq.~(\ref{bigr2}) for \( R_2 \) are either all ``$+$" or all ``$-$".
Using the same method as above, we directly find that both of these two extensions also diverge in the high-temperature limit, even for $\tilde x\neq x$.

\subsection{Free Particle and Harmonic Oscillator}

Now we illustrate the divergence of the extension $R(\tilde x,x,\tau)$ of the HK representation, which is given by Eq.~(\ref{bigr}),
for two specific examples, i.e., a free particle and a harmonic oscillator with angular frequency $\omega$. For these two examples the potential energy is $V(q)=0$ and $V(q)=m\omega^2q^2/2$, respectively. Accordingly, the trajectories 
$q_\tau$ and $p_\tau$ of Eq.~(\ref{hef2}) can be expressed as
 \begin{eqnarray}
  q_{\tau}=q+\frac{p}{m}\tau,&\ \ \ &p_{\tau}=p,\hspace{4.4cm}{\rm (free\ particle)};
  \label{freea}\\[8pt]
  q_{\tau}=
  \frac{p}{m\omega}{\sinh}(\omega\tau) +q{\cosh}(\omega\tau),
  &\ \ \ &p_{\tau}=  p{\cosh}(\omega\tau) +qm\omega{\sinh}(\omega\tau),
\ \ \ \ \ {\rm (harmonic\ oscillator).}
  \label{hoa}
   \end{eqnarray}
Substituting these results into Eqs.~(\ref{bigr}, \ref{bigr1}, \ref{bigr2}), we find that $R_1(q,p,\tau)$ is 
independent of $(q,p)$, and $R_2(\tilde x, x,q,p,\tau)$
is a Gaussian function of $q$ and  $p$.
Therefore, we have \( R(\tilde{x}, x, \tau) = R_1 \int dp \, dq \, R_2(\tilde{x}, x, q, p, \tau) \), and the integral \( \int dp \, dq \, R_2(\tilde{x}, x, q, p, \tau) \) is a Gaussian integral. 
It is straightforward to determine whether this integral diverges by examining the positive definiteness of the coefficient matrix of the quadratic terms of \( p \) and \( q \) in the exponent of \( R_2(\tilde{x}, x, q, p, \tau) \).
 Specifically, we find that no matter if $\tilde x=x$ or $\tilde x\neq x$, we always have
\begin{eqnarray}
R(\tilde x,x,\tau)=\infty,\hspace{0.4cm}{\rm for}
\left\{
\begin{array}{lll}
\tau<\frac{m}{\gamma},&&{\rm (free\ particle);}\\[12pt]
\tau<\frac{\ln 9}2\frac{m}{\gamma}\approx1.10\frac{m}{\gamma},&&{\rm (harmonic\ oscillator).}\\
\end{array}
\right.
\end{eqnarray}
Therefore,  $R(\tilde x,x,\tau)$ of a free particle or a harmonic oscillator diverges for $\tau\rightarrow 0$ ($T\rightarrow \infty$), even for 
$\tilde x\neq x$.

\section{Details of the Derivation of Eq.~(\ref{asu2}) for $N=1$}
\label{d1b}

In this appendix we show the details of the derivation of Eq.~(\ref{asu2}) for a 1D single-particle system.
As mentioned in Sec.~\ref{de},
we express the matrix element of the Boltzmann operator as
 \begin{eqnarray}
 K_\tau(\tilde x,x)=A\int dpdq\left[ F_D e^{-\frac{(S_\tau+B_\tau+C_\tau)}{\hbar}}\right],
 \label{tapp}
 \end{eqnarray}
where the factors $A$, $S_\tau$, $B_\tau$ and $C_\tau$ are given by Eqs. (\ref{cast}), (\ref{s}), (\ref{biga}) and (\ref{bigb}), respectively, for $N=1$, i.e., we have
\begin{eqnarray}
A&=&\left(\frac{\gamma}{2\hbar^{3}\pi^{3}}\right)^{1/2};\label{bigaa1}\\
S_\tau&=&\int_{0}^{\tau}d\eta \bigg[ \frac{{p}_{\eta}^{2}}{2m}+V(q_{\eta})\bigg];\label{bigsa}\\
B_\tau & = & \gamma\left(x-q\right)^{2}+\gamma\left(\tilde{x}-{q}_{\tau}\right)^{2}-p\left(x-q\right)+p_{\tau}\left(\tilde{x}-q_{\tau}\right); \label{bigaa}
\\C_\tau & = & \frac{m}{\tau}\left[\left(\tilde{x}-q_{\tau}\right)-\left(x-q\right)\right]^{2}.\label{bigba}
\end{eqnarray}
Here  $ p_{\eta}$ and $q_{\eta}$ ($0\leq \eta\leq\tau$)
satisfy the Hamilton's equations and initial condition:
\begin{eqnarray}
\frac{d}{d\eta}q_{\eta} & = & \frac{ p_{\eta}}{m};\hspace{1cm}
\frac{d}{d\eta} p_{\eta} =  \frac{d V(z)}{dz}\bigg\vert_{z=q_\eta},\label{h1a}\\
q_{\eta=0} & = & q;\hspace{1.4cm}   p_{\eta=0}=p.\label{ica}
\end{eqnarray}
Furthermore, in Eq.~(\ref{tapp})
$F_D$ is a to-be-determined function of $(\tau,p,q)$.

\subsection{Proof of Eq.~(\ref{int2})}
\label{pic}

 We first consider the system with potential $V(q)$ being of type (a) of Sec.~I, and calculate the integral $A\int dpdq \  e^{-\frac{(S_\tau+B_\tau+C_\tau)}{\hbar}}$ for $\tau\rightarrow 0$.
Let us first perform the integration for $p$. 
Since $|dV(q)/dq|$ has a finite upper bound in the total real space, in the limit  $\tau\rightarrow 0$
  the solution of the Hamilton's equations (\ref{h1a}-\ref{ica})   converges to the one of free motion, i.e., $q_\eta=q+\eta p/m$ and $p_\eta=p$. We substitute this solution into the definitions (\ref{bigsa}, \ref{bigaa}, \ref{bigba}) of $S_\tau$, $B_\tau$, and $C_\tau$, and then obtain
\begin{eqnarray}
A\int _{-\infty}^{+\infty}dp \  e^{-\frac{(S_\tau+B_\tau+C_\tau)}{\hbar}}=F_a(\tau)e^{-\frac{F_b(\tau,q)}{\hbar}},
\label{b8app}
\end{eqnarray}
where 
\begin{eqnarray}
F_a(\tau)&=&\frac{m}{\hbar\pi}\sqrt{\frac{\gamma}{\tau(m+2\tau\gamma)}};\\
F_b(\tau,q)&=&\frac{m^2 (x - \tilde x)^2 + 
 2 m \tau (2 q^2 - 4 q x + 3 x^2 - 2 x \tilde x + {\tilde x}^2) \gamma + 
 4 \tau^2 (q - x)^2 \gamma^2}{2 \tau (m + 2 \tau \gamma)}
 +\int_0^\tau V\left(q+\frac{\eta p}{m}\right)d\eta.
 \end{eqnarray}
Thus, we have
\begin{eqnarray}
\lim_{\tau\rightarrow 0}A\int dpdq \  e^{-\frac{(S_\tau+B_\tau+C_\tau)}{\hbar}}=
\lim_{\tau\rightarrow 0}\int_{-\infty}^\infty dq F_a(\tau)e^{-\frac{F_b(\tau,q)}{\hbar}}.\label{inta}
 \end{eqnarray}
Furthermore, in the limit $\tau\rightarrow 0$, we can expand 
 $F_a$ and $F_b$ in the integrand of Eq.~(\ref{inta}) as powers of $\tau$, and
ignore the terms proportional to $\tau^n$ ($n> 0$). This approach leads to
\begin{eqnarray}
\lim_{\tau\rightarrow 0}A\int dpdq \  e^{-\frac{(S_\tau+B_\tau+C_\tau)}{\hbar}}
= 
\frac{1}{\hbar\pi}\sqrt{\frac{m\gamma}{\tau}}\lim_{\tau\rightarrow 0} 
\int_{-\infty}^\infty dq
e^{-\frac{1}{\hbar}\left[\frac{m\left(x - \tilde x\right)^2}{2\tau} + 
 2\left(q^2 \gamma- 2 q x\gamma + x^2 \gamma\right)\right]}.\label{inta2}
 \end{eqnarray}
Performing the integration in the r. h. s. of Eq.~(\ref{inta2}) directly, 
 we obtain 
 \begin{eqnarray}
\lim_{\tau\rightarrow 0}A\int dpdq \  e^{-\frac{(S_\tau+B_\tau+C_\tau)}{\hbar}}
&=&\lim_{\tau\rightarrow 0}\left({\frac{m}{2\pi\hbar\tau}}\right)^{1/2}
e^{-\frac{m(x-\tilde x)^2}{2\hbar\tau}}.\label{inta1}
\end{eqnarray}
Additionally, substituting the result $\left({\frac{m}{2\pi\hbar\tau}}\right)^{1/2}
e^{-\frac{m(x-\tilde x)^2}{2\hbar\tau}}=\langle \tilde x |
e^{
-\frac{{\hat p}^2}{2m}\tau
}|x\rangle$ 
into Eq.~(\ref{inta1}), we further obtain
\begin{eqnarray}
\lim_{\tau\rightarrow 0}A\int dpdq \  e^{-\frac{(S_\tau+B_\tau+C_\tau)}{\hbar}}&=&\lim_{\tau\rightarrow 0}
\langle \tilde x |
e^{
-\frac{{\hat p}^2}{2m}\tau
}|x\rangle
=\delta(x-\tilde x),\label{int2a}
\end{eqnarray}
which is just Eq.~(\ref{int2}) of Sec.~\ref{de}.

 Now we consider the systems with  $V(q)$ bring of type-(b) as described in Sec.~I, i.e., the cases that
$V(q)$ can be expressed as $V(q)=\tilde V(q)+m\omega^2q^2/2$, and $|d\tilde V(q)/dq|$ has a finite upper bound.
For this case we can also prove  Eq.~(\ref{int2}) with the above approach. Specifically, in this case   the solution of the Hamilton's equations (\ref{h1a}-\ref{ica})   converges to the one of a harmonic oscillator, i.e., $q_{\tau}=
  \frac{p}{m\omega}{\sinh}(\omega\tau) +q{\cosh}(\omega\tau)$ and $p_{\tau}=  p{\cosh}(\omega\tau) +qm\omega{\sinh}(\omega\tau)$, for the limit  $\tau\rightarrow 0$. As in the last subsection, substituting this solution into the definitions (\ref{bigsa}, \ref{bigaa}, \ref{bigba}) of $S_\tau$, $B_\tau$, and $C_\tau$, and  performing the calculations beginning from the left-hand-side of Eq.~(\ref{b8app}), we can finally derive Eq.~(\ref{int2a}), i.e., Eq.~(\ref{int2}), again.

\subsection{Derivation of Eq.~(\ref{eqf})}
\label{proof}

\subsubsection*{1. Preliminary Calculations}

In the following we derive the equation of $F_D$, i.e., Eq.~(\ref{eqf}) of Sec. \ref{de}.
We begin from Eq.~(\ref{seqa1}), i.e.,
\begin{eqnarray}
 \hat{\Lambda}\Big[\int dpdq\left( F_D e^{-\frac{(S_\tau+B_\tau+C_\tau)}{\hbar}}\right)\Big]=0.\label{seqa}
\end{eqnarray}
For our system with $N=1$, the factors $S_\tau$, $B_\tau$ and $C_\tau$ are given by Eqs.~(\ref{bigsa}), (\ref{bigaa}) and Eq.~(\ref{bigba}), respectively. Additionally,
 the correction operator $\hat{\Lambda}$ is given by Eq.~(\ref{lam2}), i.e.,
 \begin{eqnarray}
 \hat{\Lambda}= 
 \hbar\frac{\partial}{\partial\tau}-\frac{\hbar^{2}}{2m}\frac{\partial^{2}}{\partial\tilde{x}^{2}}+V(\tilde{x})
 \label{lam2a}.
 \end{eqnarray}
 Direct calculation yields 
 \begin{eqnarray}
 \hat{\Lambda}\Big[\int dpdq\left( F_D e^{-\frac{(S_\tau+B_\tau+C_\tau)}{\hbar}}\right)\Big]=\int dpdq\left\{ F_D e^{-\frac{(S_\tau+B_\tau+C_\tau)}{\hbar}}\left[J_\tau+\hbar\frac{\dot{F_D}}{F_D}-\dot{S}_\tau+V(\tilde{{x}})\right]\right\} ,\label{c17}
\end{eqnarray}
where the dot above symbols means
\begin{eqnarray}
\dot{(...)}\equiv \frac{\partial (...)}{\partial\tau},
\end{eqnarray}
 and 
\begin{eqnarray}
J_\tau & = & \hbar\left(\frac{1}{\tau}+\frac{\gamma}{m}\right)-\frac{ p_{\tau}^{2}}{2m}+\frac{2}{m} p_{\tau}\left\{ \frac{m\left[(x-q)-(\tilde{x}-q_{\tau})\right]}{\tau}-\gamma(\tilde{x}-q_{\tau})\right\} \nonumber \\
 &  & -\frac{2}{m}\left\{ \frac{m\left[(x-q)-(\tilde{x}-q_{\tau})\right]}{\tau}-\gamma(\tilde{x}-q_{\tau})\right\} ^{2}+\frac{m}{\tau^{2}}\left[(x-q)-(\tilde{x}-q_{\tau})\right]^{2}-(\tilde{x}-q_{\tau})\dot{ p}_{\tau}\nonumber \\
 &  & + p_{\tau}\dot{q}_{\tau}+2\gamma(\tilde{x}-q_{\tau})\dot{q}_{\tau}-\frac{2m}{\tau}\left[(x-q)-(\tilde{x}-q_{\tau})\right]\dot{q}_{\tau}.\label{bigj}
\end{eqnarray}
For the convenience of the following calculations, we define
\begin{eqnarray}
u & = & \tilde{x}-q_{\tau};\label{du}\\
v & = & x-q,\label{dv}
 \end{eqnarray}
 and
\begin{eqnarray}
\Phi_\tau&=&-(B_\tau+C_\tau+S_\tau);\label{phitau}\\
V^{(n)}(z)&=&\frac{d^{n}}{dz^{n}}V(z).
 \end{eqnarray}
Substituting
 Eqs.~(\ref{h1a},  \ref{bigsa}) into Eqs.~(\ref{c17}, \ref{bigj}), and 
using the above definitions and the fact
 \begin{equation}
V(\tilde x)=V(q_\tau)+\sum_{n=1}^{\infty}u^n V^{(n)}(q_\tau),
\end{equation}
we find that Eq.~(\ref{c17}) can be re-expressed as
\begin{eqnarray}
 &  & 
 \hat{\Lambda}\Big[\int dpdq\left( F_D e^{-\frac{(S_\tau+B_\tau+C_\tau)}{\hbar}}\right)\Big]
 = \int dpdq\left\{ F_D \left[G_\tau+\hbar\frac{\dot{F_D }}{F_D }+\sum_{s=2}^{\infty}\frac{1}{s!}u^{s}V^{(s)}(q_{\tau})\right]e^{\Phi_\tau/\hbar}\right\} ,\label{res1}
\end{eqnarray}
where 
\begin{equation}
G_\tau=\hbar\left(\frac{1}{\tau}+\frac{\gamma}{m}\right)-\frac{2}{m}\left[\frac{m\left(v-u\right)}{\tau}-\gamma u\right]^{2}+\frac{m}{\tau^{2}}\left(u-v\right)^{2}.
\end{equation}

\subsubsection*{2. Eliminating $u$ and $v$}

Now we eliminate the factors $u$ and $v$ in Eq.~(\ref{res1}).
Using the fact 
\begin{eqnarray}
\frac{\partial S_\tau}{\partial q} & = & -p+\frac{\partial q_{\tau}}{\partial q} p_{\tau};\hspace{1cm}
\frac{\partial S_\tau}{\partial p}  =  \frac{\partial q_{\tau}}{\partial p} p_{\tau},
\end{eqnarray}
we find that that the factor $\Phi_\tau$ defined in Eq.~(\ref{phitau}) satisfies
\begin{equation}
\left(\begin{array}{c}
\frac{\partial\Phi_\tau}{\partial q}\\
\\
\frac{\partial\Phi_\tau}{\partial p}
\end{array}\right)
=\left(\begin{array}{cc}
{\rm R}^{qu} & {\rm R}^{qv}\\
{\rm R}^{pu} & {\rm R}^{pv}
\end{array}\right)
\left(\begin{array}{c}
u\\
\\
v
\end{array}\right).
\label{pru}
\end{equation}
Here the parameters ${\rm R}^{qu}$, ${\rm R}^{qv}$, ${\rm R}^{pu}$ and ${\rm R}^{pv}$ are given by
\begin{eqnarray}
{\rm R}^{qu} & = & -\frac{2}{\tau}m+\left(2\gamma+\frac{2}{\tau}m\right)\frac{\partial q_{\tau}}{\partial q}-\frac{\partial p_{\tau}}{\partial q};\label{r1}\\
{\rm R}^{qv} & = & 2\gamma-\frac{2}{\tau}m\left(\frac{\partial q_{\tau}}{\partial q}-1\right);\\
{\rm R}^{pu} & = & \left(2\gamma+\frac{2}{\tau}m\right)\frac{\partial q_{\tau}}{\partial p_{}}-\frac{\partial p_{\tau}}{\partial p};\\
{\rm R}^{pv} & = & 1-\frac{2}{\tau}m\frac{\partial q_{\tau}}{\partial p_{}}.\label{r4}
\end{eqnarray}
We further define  another four parameters  ${\rm T}^{qu}$, ${\rm T}^{qv}$, ${\rm T}^{pu}$ and ${\rm T}^{pv}$ via the relation:
\begin{equation}
\left(\begin{array}{cc}
{ \rm T}^{uq} & { \rm T}^{up}\\
{ \rm T}^{vq} & { \rm T}^{vp}
\end{array}\right)
=
\left(\begin{array}{cc}
{\rm R}^{qu} & {\rm R}^{qv}\\
{\rm R}^{pu} & {\rm R}^{pv}
\end{array}\right)^{-1}.
\label{uevapp}
\end{equation}
Thus, Eq.~(\ref{pru}) yields that
\begin{equation}
\left(\begin{array}{c}
u\\
\\
v
\end{array}\right)=
\left(\begin{array}{cc}
{ \rm T}^{uq} & { \rm T}^{up}\\
{ \rm T}^{vq} & { \rm T}^{vp}
\end{array}\right)
\left(\begin{array}{c}
\frac{\partial\Phi_\tau}{\partial q}\\
\\
\frac{\partial\Phi_\tau}{\partial p}
\end{array}\right).
\label{uve}
\end{equation}
Note that these R- and T-parameters are simply the R-matrices and T-matrices defined in Eqs.~(\ref{dr1}-\ref{dt}) of Sec.~(\ref{res}), with $N=1$, respectively.

For convenience of the following calculations, we introduce differential
operators $\mathbb{D}_{u}$ and $\mathbb{D}_{v}$:
\begin{eqnarray}
\mathbb{D}_{u} & = & { \rm T}^{uq}\frac{\partial}{\partial q}+{ \rm T}^{up}\frac{\partial}{\partial p};\hspace{1cm}
\mathbb{D}_{v}  =  { \rm T}^{vq}\frac{\partial}{\partial q}+{ \rm T}^{vp}\frac{\partial}{\partial p},
\end{eqnarray}
as well as the parameters 
\begin{eqnarray}
Q_{\alpha\beta} & = &\mathbb{D}_{\alpha}\left[\beta\right];\hspace{1cm}(\alpha,\beta=u,v).\label{dq}
\end{eqnarray}
Substituting the definitions (\ref{du}, \ref{dv}) of $u$ and $v$ into Eq.~(\ref{dq}), we further obtain the expressions of the $Q$-parameters:
\begin{eqnarray}
Q_{uu} & = & -{ \rm T}^{uq}\frac{\partial q_{\tau}}{\partial q}-{ \rm T}^{up}\frac{\partial q_{\tau}}{\partial p};\hspace{1cm}
Q_{uv} =  -{ \rm T}^{uq};\label{qapp1}\\
Q_{vu} & = & -{ \rm T}^{vq}\frac{\partial q_{\tau}}{\partial q}-{ \rm T}^{vp}\frac{\partial q_{\tau}}{\partial p};\hspace{1cm}
Q_{vv}  =  -{ \rm T}^{vq}.\label{qapp2}
\end{eqnarray}

Using the differential
operators $\mathbb{D}_{u}$ and $\mathbb{D}_{v}$, we can re-express Eq.~(\ref{uve}) as: 
\begin{eqnarray}
\alpha e^{\Phi_\tau/\hbar} & = & \hbar\mathbb{D}_{\alpha}\left[e^{\Phi_\tau/\hbar}\right];\hspace{1cm}(\alpha=u,v).\label{ruve}
\end{eqnarray}
Moreover, Eq.~(\ref{ruve}) leads to 
\begin{eqnarray}
\alpha\beta  e^{\Phi_\tau/\hbar}&=&\hbar\alpha\mathbb{D}_{\beta}\left[e^{\Phi_\tau/\hbar}\right]\nonumber\\
&=&
\hbar\mathbb{D}_{\beta}\left[\alpha e^{\Phi_\tau/\hbar}\right]-\hbar e^{\Phi_\tau/\hbar}\mathbb{D}_{\beta}\left[\alpha \right],
\hspace{1cm} (\alpha,\beta=u,v).\label{ee1}
\end{eqnarray}
Substituting Eqs.~(\ref{ruve}) and (\ref{dq}) into Eq.~(\ref{ee1}), we further obtain
\begin{eqnarray}
\alpha\beta  e^{\Phi_\tau/\hbar}=\hbar^2\mathbb{D}_{\beta}\left\{\mathbb{D}_{\alpha}\left[ e^{\Phi_\tau/\hbar}\right]\right\}-\hbar Q_{\beta\alpha}e^{\Phi_\tau/\hbar},
\hspace{1cm} (\alpha,\beta=u,v).\label{ee2}
\end{eqnarray}
i.e.,
\begin{eqnarray}
u^2e^{\Phi_\tau/\hbar}&=&\hbar^{2}\mathbb{D}_{u}^2\left[e^{\Phi_\tau/\hbar}\right]-\hbar Q_{uu}e^{\Phi_\tau/\hbar},\\
v^2e^{\Phi_\tau/\hbar}&=&\hbar^{2}\mathbb{D}_{v}^2\left[e^{\Phi_\tau/\hbar}\right]-\hbar Q_{vv}e^{\Phi_\tau/\hbar},\\
uve^{\Phi_\tau/\hbar}&=&\hbar^{2}\mathbb{D}_{v}\mathbb{D}_{u}\left[e^{\Phi_\tau/\hbar}\right]-\hbar Q_{vu}e^{\Phi_\tau/\hbar}.
\end{eqnarray}
Repeating  this technique, we can express any term of the form \( u^m v^n e^{\Phi_\tau/\hbar} \) ($m,n=1,2,...$) as a series in \( \hbar \), with each coefficient taking the form \( C_1 C_2 \dots [e^{\Phi_\tau/\hbar}] \), where each \( C_1, C_2, \dots \) is either a \( Q \)-factor or a \( \mathbb{D} \)-operator.
Using this approach and Eqs.~(\ref{qapp1}, \ref{qapp2}), we
can re-express Eq.~(\ref{res1}) as
\begin{eqnarray}
 \hat{\Lambda}\Big[\int dpdq\left( F_D e^{-\frac{(S_\tau+B_\tau+C_\tau)}{\hbar}}\right)\Big]
 = \hbar\int dpdq\left\{ F_D \left[\hat{L}_{\tau}+\frac{\dot{F_D }}{F_D }\right]e^{\Phi_\tau/\hbar}\right\} ,\label{res1a}
\end{eqnarray}
where 
\begin{equation}
\hat{L}_{\tau}=\hat{L}_{\tau}^{(0)}+\hbar\hat{L}_{\tau}^{(1)}+\hbar^{2}\hat{L}_{\tau}^{(2)}+...\label{bigla}
\end{equation}
Here $\hat{L}_{\tau}^{(0,1,2,...)}$ are $\hbar$-independent operators. Specially, we have
\begin{eqnarray}
\hat{L}_{\tau}^{(0)}= g_\tau(q,p),
\end{eqnarray}
where
\begin{eqnarray}
 g_\tau(q,p)\equiv\left(\frac{1}{\tau}+\frac{\gamma}{m}\right)+\left(\frac{2\gamma^{2}}{m}+\frac{m}{\tau^{2}}+\frac{4\gamma}{\tau}\right)Q_{uu}+\left(\frac{2m}{\tau^{2}}+\frac{4\gamma}{\tau}\right){\rm T}^{uq}-\frac{m}{\tau^{2}}{\rm T}^{vq}-\frac{1}{2}Q_{uu}V^{(2)}(q_{\tau}),\label{fgqp}
\end{eqnarray}
 and
\begin{eqnarray}
\hat{L}_{\tau}^{(1)}&=&-\left(\frac{2\gamma^{2}}{m}+\frac{m}{\tau^{2}}+\frac{4\gamma}{\tau}
-\frac{1}{2}V^{(2)}(q_{\tau})
\right)\mathbb D_u^2
+\left(\frac{2m}{\tau^{2}}+\frac{4\gamma}{\tau}\right)\mathbb D_u \mathbb D_v
-\frac{m}{\tau^{2}}\mathbb D_v^2
\nonumber\\
&&
-\frac{1}{3!}V^{(3)}(q_{\tau})
\bigg(
\mathbb{D}_{u}Q_{uu} 
+2Q_{uu} \mathbb{D}_{u}
\bigg)+\frac 3{4!}V^{(4)}(q_{\tau})Q_{uu}^2.
\end{eqnarray}
Notice that $g_\tau(q,p)$  is just the function given by Eq.~(\ref{ftau}) of Sec.~\ref{res}, with $N=1$.

\subsubsection*{3. Equation of $F_D $}

As in Ref.~\cite{Kay}, by repeated integration by parts and using the fact that the integrated terms tend to zero for $q\rightarrow \pm\infty$ or $p\rightarrow \pm\infty$, we can further re-express
Eq.~(\ref{res1}) as
\begin{eqnarray}
 \hat{\Lambda}\Big[\int dpdq\left( F_D e^{-\frac{(S_\tau+B_\tau+C_\tau)}{\hbar}}\right)\Big]
 = \hbar\int dpdq\left( e^{\Phi_\tau/\hbar} \hat{\tilde L}_{\tau}[F_D ]\right) ,\label{res1a}
\end{eqnarray}
where 
\begin{equation}
\hat{\tilde L}_{\tau}=\hat{\tilde L}_{\tau}^{(0)}+\hbar\hat{\tilde L}_{\tau}^{(1)}+\hbar^{2}\hat{\tilde L}_{\tau}^{(2)}+....\label{biglta}
\end{equation}
Here $\hat{\tilde L}_{\tau}^{(0,1,2,...)}$ are also a group of $\hbar$-independent operators. For instance, we have
\begin{eqnarray}
\hat{\tilde L}_{\tau}^{(0)}&=&\frac{\partial}{\partial \tau}+\hat{L}_{\tau}^{(0)}=\frac{\partial}{\partial \tau}+g_\tau(q,p),
\end{eqnarray}
and
\begin{eqnarray}
\hat{\tilde L}_{\tau}^{(1)}&=&-\tilde{\mathbb  D}_u^2\left(\frac{2\gamma^{2}}{m}+\frac{m}{\tau^{2}}+\frac{4\gamma}{\tau}
-\frac{1}{2}V^{(2)}(q_{\tau})
\right)
+\tilde{\mathbb  D}_v \tilde{\mathbb  D}_u
\left(\frac{2m}{\tau^{2}}+\frac{4\gamma}{\tau}\right)
-\tilde{\mathbb  D}_v^2 \frac{m}{\tau^{2}}
\nonumber\\
&&
-\frac{1}{3!}
\bigg(
Q_{uu}  \tilde{\mathbb  D}_{u}
+2  \tilde{\mathbb  D}_{u}Q_{uu}
\bigg)V^{(3)}(q_{\tau})
+\frac 3{4!}V^{(4)}(q_{\tau})Q_{uu}^2,
\end{eqnarray}
with
$\tilde{\mathbb  D}_{u}$ and $\tilde{\mathbb  D}_{v}$ being defined as
\begin{eqnarray}
\tilde{\mathbb  D}_{u}[...] & = & \frac{\partial}{\partial q}[{ \rm T}^{uq}...]+\frac{\partial}{\partial p}[{ \rm T}^{up}...];\hspace{1cm}
\tilde{\mathbb  D}_{v}[...]  =  \frac{\partial}{\partial q}[{ \rm T}^{vq}...]+\frac{\partial}{\partial p}[{ \rm T}^{vp}...].
\end{eqnarray}
Substituting Eq.~(\ref{res1a}) and (\ref{biglta}) into Eq.~(\ref{seqa}), 
we finally obtain Eq.~(\ref{eqf}) of of Sec. \ref{de}, i.e.,
 \begin{eqnarray}
\bigg(\hat {\tilde L}_0+\hbar \hat {\tilde L}_1+\hbar^2\hat {\tilde L}_2+...\bigg)F_D =0.\label{eqfa}
 \end{eqnarray}

\subsubsection*{4. Derivation of Eqs.~(\ref{bigcp2}, \ref{gtilde22})}
\label{fddiv}

In the end of this subsection, 
we consider the case that
if for specific $q$, $p$ (denoted as $q_\ast$, $p_\ast$),
 $g_{\eta}(q_\ast,p_\ast)$
diverges when $\eta=\eta_1,\eta_2,...,\eta_n$ ($\eta_{1,...,{n}}\in (0,\tau)$). 
We will show that in this case $F_D(q_\ast,p_\ast,\tau)$ is given by Eq.~(\ref{bigcp2}), i.e.,
 \begin{eqnarray}
 F_D(q_\ast,p_\ast,\tau)= \cos\left(\pi \tilde g\right) \cdot
 e^{-{\cal P}\int_{0}^{\tau}g_{\eta}
 (q_\ast,p_\ast)
 d\eta},
\label{bigcp2app}
 \end{eqnarray}
 with $\tilde g$ being given by Eq.~(\ref{gtilde22}), i.e.,
 \begin{eqnarray}
\tilde g= \sum_{\xi=1}^{n}\lim_{\eta\rightarrow\eta_\xi}\bigg[(\eta-\eta_\xi) \cdot g_{\eta}
 (q_\ast,p_\ast)\bigg].
 \label{gtilde222}
 \end{eqnarray}

\bigskip
\hspace{7cm}{\bf   Case with $n=1$.} 
\bigskip

We first consider the case with $n=1$, i.e., $g_{\eta}(q_\ast,p_\ast)$
diverges only when $\eta=\eta_1$ ($\eta_{1}\in (0,\tau)$). 
We introduce another group of semi-classical IVR 
 $K^{(\epsilon)}_\tau(\tilde x,x)$ ($\epsilon\in{\rm Reals}$)
 of the Boltzmann operator, by replacing the real parameter $\gamma$ of $K_\tau(\tilde x,x)$ in Eq.~(\ref{tapp})
 with the complex one $\gamma+i\epsilon$, i.e., 
 \begin{eqnarray}
 K^{(\epsilon)}_\tau(\tilde x,x)=A^{(\epsilon)}\int dpdq\left[ F_D^{(\epsilon)} e^{-\frac{(S_\tau+B_\tau^{(\epsilon)}+C_\tau)}{\hbar}}\right],\ \ \ \ \epsilon\in{\rm Reals},
 \label{tapp1}
 \end{eqnarray}
where the factors $S_\tau$ and $C_\tau$ are still given by Eqs. (\ref{bigsa}) and (\ref{bigba}), respectively, and
\begin{eqnarray}
A^{(\epsilon)}&=&\left(\frac{\gamma+i\epsilon}{2\hbar^{3}\pi^{3}}\right)^{1/2};\label{bigaa11}\\
B_\tau^{(\epsilon)} & = & (\gamma+i\epsilon)\left(x-q\right)^{2}+(\gamma+i\epsilon)\left(\tilde{x}-{q}_{\tau}\right)^{2}-p\left(x-q\right)+p_{\tau}\left(\tilde{x}-q_{\tau}\right). \label{bigaa1}
\end{eqnarray}
Additionally, using the derivation of the above subsections, it can be proven that 
\begin{eqnarray}
F_D^{(\epsilon)}(q,p,\tau)= e^{-\int_{0}^{\tau}g^{(\epsilon)}_{\eta}
 (q,p)
 d\eta},\label{feapp}
\end{eqnarray}
where $g^{(\epsilon)}_{\eta}(q,p)$ is obtained by 
replacing $\gamma$ of the expression (\ref{fgqp})  with $\gamma+i\epsilon$. Note that the right-hand-side of
Eq.~(\ref{fgqp})
includes factors $Q_{uu}$, $T^{uq}$ and $T^{vq}$. According to Eqs.~(\ref{r1}-\ref{r4}, \ref{uevapp}, \ref{qapp1}, \ref{qapp2}), these factors  are also functions of \( \gamma \). In the above derivation for $g^{(\epsilon)}_{\eta}(q_\ast,p_\ast)$, the \( \gamma \) that $Q_{uu}$, $T^{uq}$ and $T^{vq}$ depend on should also be replaced by \( \gamma+i\epsilon \).

It is clear that 
$\lim_{\epsilon\rightarrow 0^{\pm}}g^{(\epsilon)}_{\eta}
 (q,p)=g_{\eta}
 (q,p)$, with
 $g_{\eta}
 (q,p)$ being given by Eq.~(\ref{fgqp}).
  Moreover, if $g_{\eta}
 (q_\ast,p_\ast)$ diverges at  $\eta_1\in(0,\tau)$, i.e.,  
 \begin{eqnarray}
g_{\eta}(q_\ast,p_\ast)=\frac{\tilde g}{\eta-\eta_1}+{\cal O}(1),\hspace{1cm} {\rm for}\ \eta\rightarrow\eta_1,
\label{g44}
\end{eqnarray}
with $
\tilde g= \lim_{\eta\rightarrow\eta_1}\big[(\eta-\eta_1) \cdot g_{\eta}
 (q_\ast,p_\ast)\big]
 $, then for very small $|\epsilon|$ we have 
 \begin{eqnarray}
g^{(\epsilon)}_{\eta}(q_\ast,p_\ast)=\frac{\tilde g}{\eta-\eta_1+i\lambda\epsilon}+{\cal O}(1),
\hspace{1cm} {\rm for}\ \eta\rightarrow\eta_1,
\label{geps}
\end{eqnarray}
with $\lambda$ being a $\eta$-independent real parameter. 

Furthermore, Eqs.~(\ref{bigaa11}) yields that
$\lim_{\epsilon\rightarrow 0^{\pm}}A^{(\epsilon)}=A$
and
$\lim_{\epsilon\rightarrow 0^{\pm}}B_\tau^{(\epsilon)}=B_\tau$.  Using these facts,
 we can re-express $K_\tau(\tilde x,x)$ in Eq.~(\ref{tapp}) as 
\begin{eqnarray}
 K_\tau(\tilde x,x)=\frac{1}{2}\lim_{\epsilon\rightarrow 0^+}\bigg[K^{(+\epsilon)}_\tau(\tilde x,x)+K^{(-\epsilon)}_\tau(\tilde x,x)\bigg],
  \end{eqnarray}
  with
  \begin{eqnarray}
F_D(q,p,\tau)&=& \frac{1}{2}\lim_{\epsilon\rightarrow 0^+}\bigg[F_D^{(+\epsilon)}+F_D^{(-\epsilon)}\bigg].\label{feapp2}
\end{eqnarray}
Substituting Eq.~(\ref{feapp}) into Eq.~(\ref{feapp2}), we obtain
\begin{eqnarray}
F_D(q,p,\tau)&=&
\frac{1}{2}\lim_{\epsilon\rightarrow 0^+}\bigg[
e^{-\int_{0}^{\tau}g^{(+\epsilon)}_{\eta}
 (q,p)
 d\eta}
 +
 e^{-\int_{0}^{\tau}g^{(-\epsilon)}_{\eta}
 (q,p)
 d\eta}
\bigg]\nonumber\\
&=&
\frac{1}{2}\bigg\{
e^{-\left[\lim_{\epsilon\rightarrow 0^+}\int_{0}^{\tau} g^{(\epsilon)}_{\eta}
 (q,p)
 d\eta\right]}
 +
 e^{-\left[\lim_{\epsilon\rightarrow 0^-}\int_{0}^{\tau} g^{(\epsilon)}_{\eta}
 (q,p)
 d\eta\right]}
\bigg\}.\label{fdnew}
\end{eqnarray}
Now let us focus on a specific $(q,p)$, denoted as $(q_\ast,p_\ast)$. According to the aforementioned relation $\lim_{\epsilon\rightarrow 0^{\pm}}g^{(\epsilon)}_{\eta}
 (q,p)=g_{\eta}
 (q,p)$, if
$g_{\eta}
 (q_\ast,p_\ast)$ does not diverge for all $\eta\in(0,\tau)$, then Eq.~(\ref{fdnew}) yields $F_D(q_\ast,p_\ast,\tau)=e^{-\int_{0}^{\tau}g_{\eta}
 (q_\ast,p_\ast)
 d\eta},
$ which is same as Eq.~(\ref{fcr}) in our main text. If  $g^{(\epsilon)}_{\eta}
 (q_\ast,p_\ast)$ diverges at $\eta_1$, then using Eq.~(\ref{geps}) and the fact $\frac{1}{z+i0^{\pm}}={\cal P}\left(\frac 1z\right)\mp i\pi\delta(z)$, 
 as well as Eq.~(\ref{fdnew}), 
 one can directly find that $F_D(q_\ast,p_\ast,\tau)$ is given by Eq.~(\ref{bigcp2app}) or  Eq.~(\ref{bigcp2}) in our main text, with $\tilde g= \lim_{\eta\rightarrow\eta_\xi}[(\eta-\eta_1) \cdot g_{\eta}
 (q_\ast,p_\ast)]$, i.e., Eq.~(\ref{gtilde222}) or Eq.~(\ref{gtilde22}), with $n=1$.

 \bigskip
\hspace{7cm}{\bf   Cases with $n>1$.} 
\bigskip
 
For the cases with \( n > 1 \), we can prove Eqs.~(\ref{bigcp2}, \ref{gtilde22}) using the same approach as above. Note that for these cases, Eqs.~(\ref{g44}) and (\ref{geps}) should be replaced with \( g_{\eta}(q_\ast, p_\ast) = \frac{\tilde{g}^{(\xi)}}{\eta - \eta_\xi} + \mathcal{O}(1) \) for \( \eta \to \eta_\xi \), and \( g_{\eta}^{(\epsilon)}(q_\ast, p_\ast) = \frac{\tilde{g}^{(\xi)}}{\eta - \eta_\xi + i\lambda_\xi\cdot\epsilon} + \mathcal{O}(1) \) for \( \eta \to \eta_\xi \), respectively, where \( \xi = 1, \dots, n \), \( \tilde{g}^{(\xi)} = \lim_{\eta \to \eta_\xi} \left[ (\eta - \eta_\xi) \cdot g_{\eta}(q_\ast, p_\ast) \right] \), 
and $\lambda_\xi$ is a $\eta$-independent real parameter. Additionally, in the calculations for \( \lim_{\epsilon \to 0^\pm} \int_0^{\tau} g_{\eta}^{(\epsilon)}(q, p) \, d\eta \) in Eq.~(\ref{fdnew}), the formula \( \frac{1}{z + i0^{\pm}} = \mathcal{P}\left( \frac{1}{z} \right) \mp i\pi \delta(z) \) should be applied in the neighborhoods of each point \( \eta = \eta_\xi \) (\( \xi = 1,\dots, n \)).

 \section{Harmonic Oscillators}
 \label{ho}

In this appendix we show that our HK-like IVR for the   Boltzmann operator is exact for harmonic oscillators. We first consider a single harmonic oscillator with frequency $\omega$, and use the natural unit $\hbar=m=\gamma=1$.
As shown in Sec.~\ref{exa}, for this system 
we have $V=\omega^2 x^2/2$, and thus 
$q_{\eta}=c e^{\omega\eta}+de^{-\omega\eta}$ and $p_{\eta}=\omega(ce^{\omega\eta}-de^{-\omega\eta})$, where $c=(p +\omega q)/(2\omega)$ and $d=(\omega q-p)/(2\omega)$. Substituting these results into 
the definition of $g_\tau$, we find that 
\begin{eqnarray}
g_\tau=-\frac{-8 e^{\omega\tau} \omega + 
 e^{2\omega\tau} (-2 + \omega) \big[4 - 4 \omega\tau + 
    \tau^2 (-2 + \omega)\omega\big] + (2 + \omega) \big[4 + 4 \tau \omega +
     \tau^2\omega(2 + \omega)\big]}
     {2 \tau[-2 + e^{\tau \omega} (-2 + \omega) - \omega] \big\{4 + 
   e^{\tau \omega} \big[-4 + \tau (-2 + \omega)\big]+ \tau(2 + \omega)\big\}}.
\end{eqnarray}
Thus, the function $D_\tau=e^{-\int_0^\tau g_\eta d\eta}$ can be expressed as
\begin{eqnarray}
D_\tau=\frac{1}{2\sqrt{2}}\sqrt{
\frac
{e^{-\tau \omega} \big[-2 + e^{\tau \omega} (-2 + \omega) - \omega\big] \big\{4 + 
   e^{\tau \omega} \big[-4 + \tau (-2 + \omega)\big] + \tau (2 + \omega)\big\}}
{\omega\tau}
}.\label{hoe}
\end{eqnarray}
One can verify this result by substituting Eq.~(\ref{hoe}) into the equation $\frac{d}{d\tau}D_\tau=-g_\tau D_\tau$ and the condition $D_{\tau=0}=1$.

Substituting Eq.~(\ref{hoe}) and the above expressions of $q_\eta$ and $p_\eta$ into Eq.~(\ref{asu2}), we find that the integration $\int dpdq...$ is just a Gaussian integration. Perform this integration analytically, we find that in the SI, the r. h. s. of  Eq.~(\ref{asu2}) is just $
\left(\frac{m\omega}{2\pi\hbar\sinh(\omega \tau)}\right)^{1/2}e^{-\frac{m\omega\left[
\left(x^2+\tilde x^2\right)\cosh(\omega \tau)-2x\tilde x
\right]}{2\hbar\sinh(\omega \tau)}}$, i.e., the exact imaginary-time propagator of this harmonic oscillator. 
Furthermore, the above calculation can be directly generalized to the general cases with $N$ harmonic oscillators, and we can find that for these cases the r. h. s. of  Eq.~(\ref{asu2}) is Eq.~(\ref{ho3}) of Sec.~\ref{exa}.

\end{document}